\documentclass[
 reprint,
 nofootinbib,
 amsmath,
 amssymb,
 aps
]{revtex4-2}

\usepackage{graphicx}
\usepackage{dcolumn}
\usepackage{bm}
\usepackage{hyperref}
\usepackage{color}
\usepackage[dvipsnames]{xcolor}
\usepackage{orcidlink}

\makeatletter
\newcommand{\bbf}[1]{%
  \ifmmode
    \bm{#1}
  \else
    \textbf{#1}
  \fi
}
\makeatother

\newcommand{\bicocca}{Dipartimento di Fisica “G. Occhialini", Università degli Studi di Milano-Bicocca, Piazza della Scienza 3, I-20126 Milano, Italy}

\newcommand{\infnmilano}{INFN, Sezione di Milano-Bicocca, Piazza della Scienza 3, I-20126 Milano, Italy}

\begin{document}

\preprint{APS/123-QED}

\title{Modeling Non-Gaussianities in Pulsar Timing Array data analysis using Gaussian Mixture Models}

\author{Mikel Falxa~\orcidlink{0000-0002-0815-1781}}
\thanks{mikel.falxa@unimib.it}
\affiliation{\bicocca}
\affiliation{\infnmilano}

\author{Alberto Sesana~\orcidlink{0000-0003-4961-1606}}
\affiliation{\bicocca}
\affiliation{\infnmilano}
 
\date{\today}

\begin{abstract}
In Pulsar Timing Array (PTA) data analysis, noise is typically assumed to be Gaussian, and the marginalized likelihood has a well-established analytical form derived within the framework of Gaussian processes. However, this Gaussianity assumption may break down for certain classes of astrophysical and cosmological signals, particularly for a gravitational wave background (GWB) generated by a population of supermassive black hole binaries (SMBHBs). In this work, we present a new method for testing the presence of non-Gaussian features in PTA data. We go beyond the Gaussian assumption by modeling the noise or signal statistics using a Gaussian mixture model (GMM). An advantage of this approach is that the marginalization of the likelihood remains fully analytical, expressed as a linear combination of Gaussian PTA likelihoods. This makes the method straightforward to implement within existing data analysis tools. Moreover, this method extends beyond the free spectrum analysis by producing posterior probability distributions of higher-order moments inferred from the data, which can be incorporated into spectral refitting techniques. We validate the model using simulations and demonstrate the sensitivity of PTAs to non-Gaussianity by computing the Bayes factor in favor of the GMM as a function of the injected excess moments. We apply the method to a more astrophysically motivated scenario where a single SMBHB is resolved on top of a Gaussian GWB and show that significant non-Gaussianities are introduced by the individual source. Finally, we test our model on a realistic GWB generated from a simulated population of SMBHBs.

\end{abstract}

\maketitle

\section{Introduction}

Pulsar Timing Arrays (PTAs) provide a unique observational tool to explore gravitational waves (GWs) at nanohertz frequencies, far below those accessible to ground or space-based interferometers \cite{Sazhin, Detweiler, Foster_Backer}. In this low-frequency band, the most promising sources of GWs are supermassive black hole binaries (SMBHBs), which are expected to form naturally as a consequence of galaxy mergers over cosmic time \cite{Sesana_2008, Sesana_2013}. When two massive galaxies collide, their central black holes are brought together, eventually forming bound binary systems that lose energy through GW emission and spiral inward until merger. The combined GW emissions of an astrophysical population of such binaries results in a stochastic gravitational wave background (GWB). The spectral shape and amplitude of this background are directly connected to the merger history of massive galaxies and the dynamical environment of their central black holes \cite{Sesana_2013}. Additional sources of nanohertz frequency GWs may include cosmological phenomena such as first-order phase transitions, cosmic strings, or other relics from the early Universe \cite{Caprini_2018}.

The GWB produced by SMBHBs is expected to manifest as a low-frequency red noise process in pulsar timing data, spatially correlated between millisecond pulsars according to the Hellings–Downs (HD) curve when sources are isotropically distributed in the sky \cite{HD, bruce_variance}. Several PTA collaborations have recently reported strong evidence for a common-spectrum process consistent with the properties of a GWB with the expected HD spatial correlations \cite{epta_wm3, ng15yr, ppta_dr3, mpta, cpta}. However, standard PTA analysis techniques assume that the signal and noise are both Gaussian stochastic processes \cite{new_advances_gp}. This assumption is well motivated when the background is generated by a large number of unresolved sources, but may break down when only a few bright binaries contribute significantly, or when deterministic sources such as continuous gravitational waves (CGWs) are superimposed on the background \cite{laal2024deepneuralemulationsupermassive}. In these cases, the statistical distribution of the signal can deviate from Gaussianity, introducing higher-order structure such as skewness or excess kurtosis in the data.

In this work, we develop a method to incorporate non-Gaussian statistics into PTA data analysis by modeling the Fourier coefficients of the signal as drawn from a Gaussian mixture model (GMM). This approach allows for analytical marginalization of the likelihood while introducing flexibility to capture heavy tails or excess higher-order moments in the signal distribution, complementing previous studies \cite{lentati_ng}. We explore the method's performance on simulated datasets including individual pulsar noise, common red noise with excess kurtosis, and composite signals formed by the superposition of a Gaussian background and a small number of CGWs. Additionally, we apply our model to search for non-Gaussianites in a realistic signal that is generated from the incoherent sum of CGWs produced by a simulated population of SMBHBs. The method enables the inference of posterior distributions for higher-order moments (e.g., kurtosis), which can be used in spectral refitting techniques to gain additional insight into the statistical structure and astrophysical origin of the observed signals. By providing a principled extension to the Gaussian framework, this work offers a practical pathway to account for and exploit non-Gaussian features in PTA data analysis.

The paper is organized as follows. First, we present the standard Gaussian likelihood that is traditionally used in PTA data analysis. Then, we show how GMMs can model non-Gaussian behaviors and be readily incorporated in current data analysis techniques to produce posterior distributions of higher order moments. Finally, we test the model on simulated data to compare a Gaussian model with the non-Gaussian model. The latter is tested in an astrophysically motivated scenario where a CGW is resolved on top of either a purely Gaussian GWB or a realistic GWB.

\section{Method}

In Bayesian analysis, our knowledge of the parameters in a given model is updated by inferring their posterior probability distribution given some prior knowledge and the outcome of the experiment (cf. Bayes theorem \cite{bayesian}). This inference requires specifying a prior probability distribution for each parameter (i.e., a prior assumption about their behavior). For example, assuming Gaussian noise corresponds to assigning normally distributed priors to the noise parameters.

\subsection{Gaussian noise}

Consider a dataset $\delta t$ with uncorrelated diagonal Gaussian noise matrix $\bm N$. The data contains time-correlated noise that is modeled using its spectral decomposition on a discrete Fourier basis $\bm F$ with a vector of Gaussian distributed coefficients $\vec{a} \sim \mathcal{N}(0, \bm \Phi)$ where the covariance matrix $\bm \Phi$ is a function of hyper-parameters $\vec \lambda$. The posterior distribution $p(\vec{a}, \vec \lambda | \delta t)$ can be written as the product of, respectively, the Gaussian likelihood $\mathcal{L} (\delta t | \vec{a})$, the prior probability $p(\vec{a}|\vec \lambda)$ on coefficients $\vec{a}$ and the prior probability $\Pi(\vec \lambda)$ on hyper-parameters $\vec \lambda$

\begin{equation}
    \ln p(\vec{a}, \vec \lambda | \delta t) = \ln \mathcal{L} (\delta t | \vec{a}) + \ln p(\vec{a}|\vec \lambda) + \ln \Pi(\vec \lambda).
\label{eq:posterior_prob}
\end{equation}

When the assumption of Gaussian noise is enforced, we set a Gaussian prior probability distribution denoted $\Pi(\vec{a}|\vec \lambda)$. In this case, this expression becomes

\begin{equation}
\begin{aligned}
    \ln p(\vec{a}, \vec \lambda | \delta t) = &-\frac{1}{2} \left [ \delta t - \bm Fa \right ]^\top \bm N^{-1} \left [ \delta t - \bm F a \right ] -\frac{1}{2}\ln\det2\pi \bm N\\
    & -\frac{1}{2} a^\top \bm \Phi^{-1} a  -\frac{1}{2}\ln\det2\pi \bm\Phi\\
    & + \ln \Pi (\vec \lambda),
\end{aligned}
\end{equation}
with $\bm \Phi = \rm diag \{S(f_n, \vec{\lambda}) \Delta f_n \}$ and $S(f_n, \vec \lambda)$ the one-sided power spectral density (PSD), $f_n=n/T$ for $n=[1, 2, ..., N_f]$ and $\Delta f_n = f_{n+1} - f_n$. Setting $\bm \Phi$ as diagonal means assuming stationary noise (even though spectral leakage due to the finite observation time window can introduce spurious off-diagonal terms \cite{crisostomi2025}).

This expression can be marginalized with respect to the Fourier coefficients $\vec a$ (see Appendix ~\ref{app:margin_likelihood}) to give the marginalized likelihood expression commonly used in PTA data analysis \cite{new_advances_gp}

\begin{equation}
\begin{aligned}
    \ln \int p(\vec a, \vec \lambda | \delta t)d\vec a & =  -\frac{1}{2} \delta t ^\top \bm C^{-1} \delta t -\frac{1}{2}\ln\det \{ 2\pi \bm C \} + \ln \Pi (\vec \lambda)\\
    & = \ln \mathcal{L}(\delta t | \vec \lambda) + \ln \Pi (\vec \lambda),
\end{aligned}
\label{eq:margin_post}
\end{equation}
where $\bm C = \bm N + \bm F^\top \bm \Phi \bm F$ and $\mathcal{L}(\delta t | \vec \lambda)$ is the marginalized Gaussian likelihood.

In PTA specifically, the $\bm F$ matrix is the concatenation of the Fourier basis of all the noise sources present in the data with the timing model design matrix. The latter accounts for first order errors in the pulsar timing fit that is performed to obtain the timing residuals $\delta t$ \cite{ng_noise_budget}.

\subsection{Non-Gaussian noise}

\begin{figure}[h!]
\centering
    \includegraphics[width=0.45\textwidth]{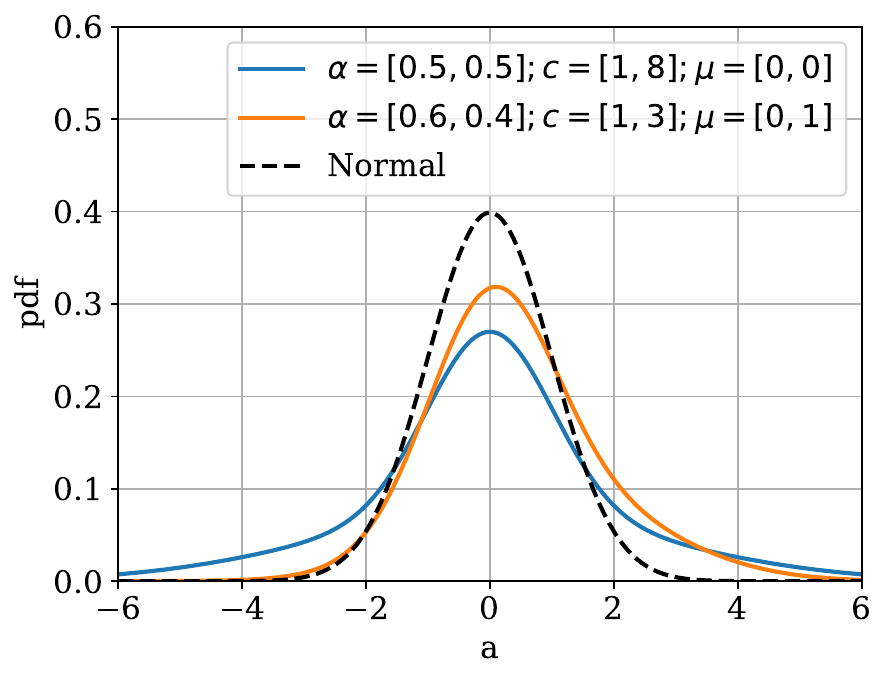}
    \caption{Example of distributions obtained with a GMM of 2 Gaussians with $\Phi=1$ (see \autoref{eq:gaussian_mixture}) and different sets of parameters $\alpha$, $\mu$ and $c$ introducing heavier tails or asymmetries (i.e. kurtosis and skewness).}
    \label{fig:example_ng}
\end{figure}

In the case of non-Gaussian noise, we can write the same posterior as in \autoref{eq:posterior_prob}, with the difference that $\vec a \sim p_a(\vec \lambda_{NG})$, meaning that the Fourier coefficients are distributed according to $p_a(\vec \lambda_{NG})$, a non-Gaussian probability distribution that is a function of hyper-parameters $\vec \lambda_{NG}$. The problem is that the marginalization cannot be performed analytically \cite{lentati_ng}. Still, we can model any probability distribution using a GMM as

\begin{equation}
\begin{aligned}
    p_a (a|\Phi,\vec{\alpha}, \vec{\mu}, \vec{c}) & = \sum_{i=1}^N \alpha_i \frac{\exp \left \{ -\frac{1}{2} \left [a - \mu_i\right ]^2 (c_i\Phi)^{-1}\right \}}{\sqrt{2\pi (c_i\Phi) }}\\
    & = \sum_{i=1}^N \alpha_i \varphi(a, \mu_i, \sqrt{c_i \Phi}),
\end{aligned}
\label{eq:gaussian_mixture}
\end{equation}
with $\sum_i \alpha_i = 1$ to ensure the normalization of the distribution. We denote as $\varphi(a, \mu_i, \sqrt{c_i \Phi})$ the probability density function of a Gaussian distribution with mean $\mu_i$ and variance $c_i \Phi$ (see \autoref{fig:example_ng}).

The advantage of writing the prior in this form is that the marginalization becomes fully analytical, as in \autoref{eq:margin_post}. We consider the class of symmetric distributions centered on zero by setting $\vec{\mu}=0$. For a single Fourier coefficient $a$, following the probability distribution in \autoref{eq:gaussian_mixture}, the marginalized likelihood is

\begin{equation}
\begin{aligned}
    \ln \int p(a, \Phi,\vec{\alpha}, \vec{c}| \delta t) d a= & \ln \sum_{i=1}^N \alpha_i \mathcal{L} (\delta t | c_i \Phi)\\
    & + \ln \Pi (\Phi) + \ln \Pi (\vec c)+ \ln \Pi (\vec \alpha),
\end{aligned}
\end{equation}
where deviations from Gaussianity are controlled by the parameters $\alpha_i$ and $c_i$. In \autoref{tab:moments} we give the first central moments for a mixture of two Gaussian distributions.

One obvious caveat is that the more Gaussian distributions we use in the mixture, the more likelihood evaluations are required, making the overall computation slower (see \autoref{sec:ng_likelihood}). In particular, if the hypothesis of uncorrelated frequencies is required across the considered spectrum, the number of likelihood evaluations grows exponentially as $\sim N^{2N_f}$. We can however check gaussianity frequency bin per frequency bin.

\begin{table}[]
    \centering
    \begin{tabular}{c|c}
        $q$ & $M_q = E \{(a - \bar{a})^q\}$ \\
        \hline
        & \\
        1 & $0$ \\
        & \\
        2 & $\mu_0^2 \alpha [1 - \alpha] + \Phi [1 + \alpha (c -1)]$\\
        & \\
        3 & $\mu_0 ^3 [\alpha (1 - \alpha)^3 - \alpha^3 ( 1 - \alpha)]$ \\
         & $+ \Phi 3(1 - \alpha) \alpha \mu_0 (c - 1)$ \\
         & \\
        4 & $\mu_0^4 [\alpha^4 (1 - \alpha) + \alpha (1 - \alpha)^4]$\\
        & $+ 6\mu_0^2 \Phi[(1 - \alpha) \alpha^2 + \alpha(1 - \alpha)^2 c]$\\
        & $+3\Phi^2 [1 + \alpha (c^2 - 1)]$
    \end{tabular}
    \caption{First 4 central moments for a mixture of two Gaussian distributions with $\vec{\alpha} = [1-\alpha, \alpha]$, $\vec{c} = [1, c]$ and $\vec{\mu} = [0, \mu_0]$ (see \autoref{eq:gaussian_mixture}). We have defined $E\{a\}=\bar{a}$. Note that for $\alpha=0$, the moments are identical to those of the standard Gaussian distribution. For more details, see Appendix ~\ref{app:moments}.}
    \label{tab:moments}
\end{table}

For $\vec{\mu}=[0, 0]$ the $2q$-th central moment has a very simple expression

\begin{equation}
   M_{2q} = (2q - 1)!! \Phi^q [1 + \alpha(c^q - 1)].
\end{equation}

In this case, only the even moments $M_{2q}$ are non-zero because the distribution is symmetric, thus $M_{2q+1}=0$.

If we note $\mathcal{M}_{q} (\sigma)$ the $q$-th moment of a centered Gaussian distribution with variance $\sigma^2$, we can quantify deviations from Gaussianity by calculating the relative excess moments

\begin{equation}
    \Delta \bar{M}_{q} = \frac{M_{q} - \mathcal{M}_{q}(\sqrt{M_2})}{\mathcal{M}_{q}(\sqrt{M_2})}.
\label{eq:excess_moments}
\end{equation}

$\Delta \bar{M}_{q}$ tells us how much the probability distribution $p(a)$ is different from a Gaussian that has the same $M_2$, i.e. the same variance. If $\Delta \bar{M}_{q} \neq 0$ for $q>2$, $p(a)$ is not shaped like a Gaussian.

\subsection{Common correlated noise : higher order statistics}

If the observed signal is non-Gaussian and of gravitational origin, it is possible to calculate its higher order correlators. In PTA, a Gaussian GWB is entirely characterized by its second moment and the associated 2-point correlation function $\chi^{(2)}_{IJ}$ between 2 pulsars is the well-known HD curve \cite{HD}, describing the overlap between their response to an isotropic GW signal. Considering $N_p$ pulsars and the $n$-th Fourier frequency, the prior probability distribution for the vector of random Fourier coefficients $\vec{a}_n$ of length $N_p$ is

\begin{equation}
    \Pi(\vec{a}_n) = \frac{\exp \{-\frac{1}{2}\sum_{IJ} a_{I,n} [\chi^{(2)}_{IJ} \Phi_n]^{-1}a_{J,n}\}}{\sqrt{\det{2\pi \bm \chi^{(2)} \Phi_n}}},
\label{eq:hd_prior}
\end{equation}
where $\bm \chi^{(2)}$ is the matrix of the correlation coefficients between $\vec{a}_n$.

This expression yields correlated noise between pulsars that is responsible for off-diagonal terms in the marginalized likelihood expression (see Appendix ~\ref{app:margin_likelihood}). In \cite{seto_2009, bruce_source_anis}, we see that higher order statistics necessitate the calculation of at least the 4-point correlation functions $\chi^{(4)}_{IJKL}$ characterizing the excess kurtosis (4th cumulant) of the noise statistics, that is

\begin{equation}
    \langle a_{I,n} a_{J,n} a_{K,n} a_{L,n} \rangle \propto \chi^{(4)}_{IJKL} h_n^{(4)},
\end{equation}
where $h_n^{(4)}$ is the level of excess kurtosis and the brackets denote the ensemble average.

However, incorporating higher order correlators can become expensive in PTA analysis, because the number of pulsars that make up the data is usually large. Consequently, the number of possible combinations and the associated numerical cost grow quickly. As a first approximation, we could ignore those inter pulsar higher order correlations $\chi^{(4)}_{IJKL}$ for $I\neq J \neq K \neq L$ and still model the non-Gaussianities using lower order statistics, because the presence of excess kurtosis still introduces effects for the $I=J=K=L$ terms. Conceptually, this is the analog of modeling a common uncorrelated noise instead of an HD correlated one. We retain the statistics of the Fourier spectrum while neglecting the higher order correlation signatures, hence losing some of the information. For example, if the signal is generated by an astrophysical population of SMBHBs, the underlying Poissonian statistics should still be detectable in the spectrum without accounting for $\chi^{(4)}_{IJKL}$ \cite{xiao_xue, Lamb_2024}.

We show in Appendix ~\ref{app:kurtosis_expansion} how we can add the contributions of higher order statistics by expanding the prior probability distribution $\Pi(\vec{a}_n)$ in terms of the Gram-Charlier A series (or similarly using the Edgeworth series). This approach being computationally less efficient, it is not further explored in this work.

\subsection{Non-Gaussian Likelihood}
\label{sec:ng_likelihood}

When noise is assumed to be Gaussian and stationary, the likelihood takes a well established form that is widely used in PTA data analysis \cite{new_advances_gp}. The noise is described by its Fourier coefficients $\vec{a}$ and the assumption of stationarity requires that different frequencies are uncorrelated. In terms of prior probability, this translates into

\begin{equation}
    p(\vec{a}) = \prod_{n=1} ^{N_f} p_n (a_n^c, \Phi^c_n, \vec{\alpha}, \vec{\mu}, \vec{c}) p_n (a_n^s, \Phi^s_n, \vec{\alpha}, \vec{\mu}, \vec{c}),
\label{eq:full_ng_prior}
\end{equation}
where $p_n(a_n)$ is the prior probability for the $n$-th coefficient and $N_f$ is the number of frequency bins. There are $2N_f$ terms in total to account for $a_n^c$ and $a_n^s$, respectively the real and imaginary components of $a_n$. We reiterate that assuming each frequency to be uncorrelated is only an approximation when we have a finite observation time that might bias the inference \cite{crisostomi2025}.

If the $p_n(a_n)$ are modeled as a GMM of $N$ Gaussian distributions, the total probability $p(\vec{a})$ becomes a sum of $N ^{2N_f}$ Gaussian distributions. Then, the evaluation of the marginalized non-Gaussian likelihood would require the evaluation of $N^{2N_f}$ Gaussian likelihoods, making the process very computationally inefficient, although fully analytical. We propose two workarounds for two limiting cases, (i) using the unmarginalized likelihood form and sampling the Fourier coefficients (ii) using the marginalized likelihood by setting non-Gaussian behavior in one single frequency bin. The first approach is practical for single pulsar noise analysis, assuming that the non-Gaussian behavior of the noise is consistent across frequencies and that enough frequencies are probed to identify deviations from Gaussianity. The second approach can only be used for common noise between pulsars as it allows one frequency bin to deviate from Gaussianity. It does however require a larger number of pulsars to pick up the non-Gaussian behavior. \textcolor{red}{}\\

\textbf{Method 1. Sampling the Fourier coefficients}\\

In the case of single pulsar noise analysis, we directly sample the coefficient $\vec a$ using the likelihood

\begin{equation}
    \mathcal{L}(\delta t | \vec{a}, \vec{\Phi}, \vec{\mu}, \vec{\alpha}, \vec{c}) = \mathcal{L}(\delta t| \vec a) p(\vec{a}|\vec{\Phi}, \vec{\mu}, \vec{\alpha}, \vec{c}).
\label{eq:spa_ng_likelihood}
\end{equation}

This method is the most straightforward as it requires no analytical marginalization. In fact, it does not even require a GMM for the prior, as long as the distribution is known \cite{lentati_ng}. The main caveat is that the dimensionality of the parameter space may be very high. Previous work showed how this type of problem can be optimized using Gibbs sampling \cite{Laal_2025}.\\

\textbf{Method 2. Per frequency marginalized non-Gaussian likelihood}\\

For the per frequency marginalized non-Gaussian likelihood, we only model non-Gaussian behavior at one frequency. Then, the prior probability distribution for the Fourier coefficients is given by \autoref{eq:full_ng_prior} for $N_f=1$, where the $p_n (a_n^c, \Phi^c_n, \vec{\alpha}, \vec{\mu}, \vec{c})$ and $p_n (a_n^s, \Phi^s_n, \vec{\alpha}, \vec{\mu}, \vec{c})$, modeled as GMMs, are given by \autoref{eq:gaussian_mixture}. The marginalization of \autoref{eq:spa_ng_likelihood} with respect to $\vec{a}$ yields

\begin{equation}
    \mathcal{L}(\delta t |\vec{\Phi}, \vec{\mu}, \vec{\alpha}, \vec{c}, \vec \lambda) = \sum_{i,j=1}^N \alpha_i \alpha_j \mathcal{L}(\delta t|c_i \Phi^c_n,c_j\Phi_n^s, \mu_i,\mu_j, \vec \lambda)
\end{equation}
where $\mathcal{L}(\delta t|c_i \Phi^c_n,c_j\Phi_n^s, \mu_i,\mu_j, \vec \lambda)$ is the marginalized likelihood derived in \autoref{app:margin_likelihood} and $\vec \lambda$ is the vector of all other parameters that are not related to the non-Gaussian signal (i.e. Gaussian prior widths, timing model parameters, individual pulsar noise). For $\mu_i=\mu_j=0$, the likelihoods in the sum are identical to the commonly used PTA likelihood \cite{new_advances_gp}. Then, the computation of the non-Gaussian likelihood can be parallelized by computing each term in the sum in parallel. Moreover, this method is very well suited for reversible jump MCMC techniques where one could dynamically add or remove Gaussians to the mixture to find their optimal number and properties. This will be left for future studies.

In this work, we will only consider the case with $\Phi_n^c=\Phi_n^s$, $\vec{\mu}=0$ and a mixture of 2 Gaussians. Using \autoref{eq:gaussian_mixture}, we construct the probability distribution for the non-Gaussian noise

\begin{equation}
\begin{aligned}
    p_n (a|\Phi_n, \alpha, c) & = (1 - \alpha) \times \varphi(a,0,\sqrt{\Phi_n}) + \alpha \times \varphi(a,0,\sqrt{c\Phi_n}),
\end{aligned}
\end{equation}
that is fully characterized by three parameters $\Phi_n, \alpha$ and $c$. This yields a likelihood of the form

\begin{equation}
\begin{aligned}
    \mathcal{L}(\delta t |\Phi^c_n, \Phi^s_n, \alpha, c, \vec \lambda) & = (1-\alpha)^2 \mathcal{L}(\delta t|\Phi_n,\Phi_n, \vec \lambda)\\
    &+\alpha^2\mathcal{L}(\delta t|c \Phi_n,c\Phi_n, \vec \lambda)\\
    &+(1-\alpha)\alpha \mathcal{L}(\delta t| \Phi_n,c\Phi_n, \vec \lambda)\\
    &+\alpha (1-\alpha)\mathcal{L}(\delta t|c\Phi_n, \Phi_n, \vec \lambda).
\end{aligned}
\label{eq:ng_likelihood}
\end{equation}

This method could be generalized to any known probability distribution $p_n(a, \vec{\lambda}_{NG})$ by first fitting a GMM to $p_n(a, \vec{\lambda}_{NG})$ and fixing the recovered parameters $\vec{\alpha},\vec{\mu}$ to their best fit values to get an analytical approximation of the marginalized non-Gaussian likelihood. If a mapping function $\vec{\lambda}_{NG} \rightarrow \vec{\alpha},\vec{\mu}$ is defined, one could infer the parameters $\vec{\lambda}_{NG}$ from the data through direct sampling or spectral refitting \cite{Lamb_2023, mitridate2023ptarcade, Quelquejay_Leclere_2023}.

\subsection{Free spectrum}

In the PTA literature, a widely used data analysis technique is the free spectrum analysis \cite{epta_wm3, ng15yr, ppta_dr3, mpta, ng_noise_budget}, that consists in estimating the RMS value $\rho$ of the noise (or signal) in the data at a given frequency bin. This estimate is directly proportional to the PSD of the stochastic processes and gives crucial information about its physics. However, this estimate always assumes that the noise in that frequency bin is Gaussian, which is not necessarily true. To produce an unbiased estimator of the PSD, we need to account for the real statistic of the Fourier bins $\vec{a}$ of that process. Given that we know the distribution of $a_n$ at frequency $n$ for both its real $a_n^c$ and imaginary parts $a_n^s$ (or equivalently cosine and sine components), the free spectrum is given by

\begin{equation}
    \rho^2_n = S(f_n)\Delta f=\frac{1}{2} \left\langle | a_n^c |^2 + |a_n^s|^2 \right\rangle,
\end{equation}
where the brackets denote the ensemble average, following the definition of the PSD for stochastic processes that is related to the variance of $a_n$.

The free spectrum is obtained from the posterior samples of parameters characterizing the probability distribution of $a_n$. For Gaussian noise, it is directly given by the posterior distribution of its standard deviation $\sqrt{\Phi_n}$, since the term in the brackets behaves as a chi-square distribution with two degrees of freedom, as $\rho_n^2 = \Phi_n \langle \chi^2_2\rangle / 2 = \Phi_n$.

When $a_n$ is described as a 2 Gaussian GMM, the result is given by \autoref{tab:moments}. For more Gaussian in the GMM, we found no closed form expression of the ensemble average for the sum of the squares. However, the result can be easily obtained numerically by sampling the GMM. For a set of non-Gaussianity parameters $\Phi_n$, $\alpha$ and $c$, given that we have inferred the posterior distribution $p(\Phi_n, \alpha,c|\delta t)$ from data $\delta t$, the probability distribution $p(a_n,\Phi_n,\vec{\alpha}, \vec{c}|\delta t)$ is given by

\begin{equation}
    p(a_n, \Phi_n, \alpha,c|\delta t) = p(a_n|\Phi_n, \alpha,c)p(\Phi_n, \alpha,c|\delta t),
\label{eq:posterior_ng}
\end{equation}
that is the prior probability defined in \autoref{eq:gaussian_mixture} weighted by the posterior probability of parameters $\Phi_n,\alpha,c$. The probability distribution of $a_n$ marginalized over the non-Gaussianity parameters can be obtained from the posterior samples of $\Phi_n,\alpha,c$. For $N_D$ samples drawn from $p(\Phi_n, \alpha,c|\delta t)$, we have

\begin{equation}
\begin{aligned}
    p(a_n|\delta t) & = \int p(a_n, \Phi_n, \alpha,c|\delta t) d\Phi_n d\alpha dc\\
    &\approx \frac{1}{N_D}\sum_{j=1}^{N_D} p(a_n|\Phi_j, \alpha_j, c_j).
\end{aligned}
\end{equation}

We can also construct posterior samples of the $2q$-th moment estimators $\rho^{2q}_{n}$ of $p(a_n|\delta t)$ as

\begin{equation}
    \rho^{2q}_{n,j} \approx \frac{1}{2N_S}\sum_{i=1}^{N_S} \left[| a_{nj,i}^c |^{2q} + | a_{nj,i}^s |^{2q} \right],
\label{eq:qth_order_fs}
\end{equation}
where the $a_{nj,i}^c, a_{nj,i}^s$ are sampled from $p(a_n,\Phi_{nj},\alpha_j,c_j|\delta t)$ and $N_s$ is the total number of samples used to numerically estimate the average for the $j$-th posterior sample of $p(\Phi_n, \alpha,c|\delta t)$. Here, we discard uneven moments because we only consider symmetric distributions. In this work, since we consider uncorrelated $a^c_n$ and $a^s_n$ modeled as a 2 Gaussian GMM, $\rho^{2}_{n,j}$ and $\rho^{4}_{n,j}$ can be directly obtained from the posterior samples of $\Phi_{nj},\alpha_j,c_j$ using the expressions in \autoref{tab:moments}.

The histogram of the recovered $\rho^{2q}_{n,j}$ corresponds to the posterior distribution of the inferred $2q$-th moments. This additional information could be used in spectral refitting techniques to constrain the statistical properties of astrophysical signals like the GWB produced by a population of SMBHBs \cite{Lamb_2024, xiao_xue, bruce_source_anis}. For $q=1$, the distribution gives the already known Gaussian free spectrum that is an estimator of the PSD.

\section{Results}
\label{sec:results}

To test the performance of the model we will study two cases : the single pulsar noise analysis and the search of non-Gaussianities at a given frequency for a common uncorrelated pulsar noise. We consider a PTA of $N_p=100$ pulsars with $T_{obs}=10$ years of observation, $N_t=500$ data points ($\approx$ 1 week cadence) and $\sigma=10^{-7}$s timing uncertainty level. The data is simulated using \texttt{fakepta}\footnote{\url{https://github.com/mfalxa/fakepta}} and the Gaussian likelihoods are computed using \texttt{ENTERPRISE}\footnote{\url{https://github.com/nanograv/enterprise}}\cite{enterprise}. The timing residuals $\delta t_a$ of pulsar $a$ can be written as

\begin{equation}
    \delta t_I = w_I + \sum_n X_{I,n} \sin(2\pi f_n t) + Y_{I,n} \cos(2\pi f_n t),
\end{equation}
with $w_I \sim \mathcal{N}(0,\sigma^2 \mathcal{I}_{N_t})$ where $\mathcal{I}_{N_t}$ is the identity matrix of rank $N_t$ and $\vec{X}_{n}, \vec{Y}_{n} \sim p(\alpha, c, S(f_n)\Delta f)$ where $S(f_n)$ is the one-sided PSD of the non-Gaussian time-correlated process and $\alpha$, $c$ the parameters characterizing the non-Gaussianities of the probability distribution $p$. The PSD of the process $S(f)$ is chosen to be a powerlaw of the form

\begin{equation}
    S(A, \gamma, f) = \frac{A^2}{12\pi^2} \left( \frac{f}{f_{yr}} \right)^{-\gamma},
\label{eq:powerlaw}
\end{equation}
with $A$ the amplitude, $\gamma$ the spectral index and $f_{yr}=1/yr$ the frequency corresponding to a one year period.

For each pulsar we consider a timing model accounting for the spin-down rate of that pulsar described by a quadratic polynomial in time and we marginalize over the first order errors in the timing model parameters \cite{new_advances_gp, ng_noise_budget}. The full likelihood for common uncorrelated noise is written as a product of $N_p$ individual pulsar likelihoods given by \autoref{eq:ng_likelihood}

\begin{equation}
     \mathcal{L}(\delta t |\Phi^c_n, \Phi^s_n, \alpha, c, \vec \lambda) = \prod_{k=1} ^{N_p} \mathcal{L}_k(\delta t_k |\Phi^c_n, \Phi^s_n, \alpha, c, \vec \lambda_k)
\end{equation}

\begin{table}[]
    \centering
    \begin{tabular}{c|c}
        $\log_{10} A$ & [-18, -11] \\
        $\gamma$ & [1, 7] \\
        $\alpha$ & [0, 1] \\
        $\log_{10} \sqrt{c}$ & [0, 2] \\
        $\log_{10} \sqrt{\Phi}$ & [-10, -4]
    \end{tabular}
    \caption{Uniform prior ranges for the parameters that are used in this work. The parameters $\log_{10} A$ and $\gamma$ are used when we assume a powerlaw spectrum. The free spectrum fits use the parameter $\log_{10} \sqrt{\Phi}$.}
    \label{tab:prior_ranges}
\end{table}

\subsection{Single pulsar noise}

We test the model by injecting a time-correlated noise that has a consistent statistic across all frequency bins. In this example, we consider a noise that has Fourier coefficients $a$ distributed accroding to GMM of 2 Gaussians with $\alpha=0.5$ and $c=3$, following a powerlaw as in \autoref{eq:powerlaw} with $\log_{10} A=-13.5$ and $\gamma=3$ across 50 frequency bins that are the 100 first harmonics of $1/T_{obs}$. Recent works have shown the advantages of directly sampling the Fourier coefficients \cite{Laal_2025, valtolina2025regularizingpulsartimingarray}. The only disadvantange comes from the high number of coefficients there are to sample, but the recovered posterior for the unmarginalized parameters are identical to the traditionally marginalized likelihood.

\begin{figure}
\centering
    \includegraphics[width=0.45\textwidth]{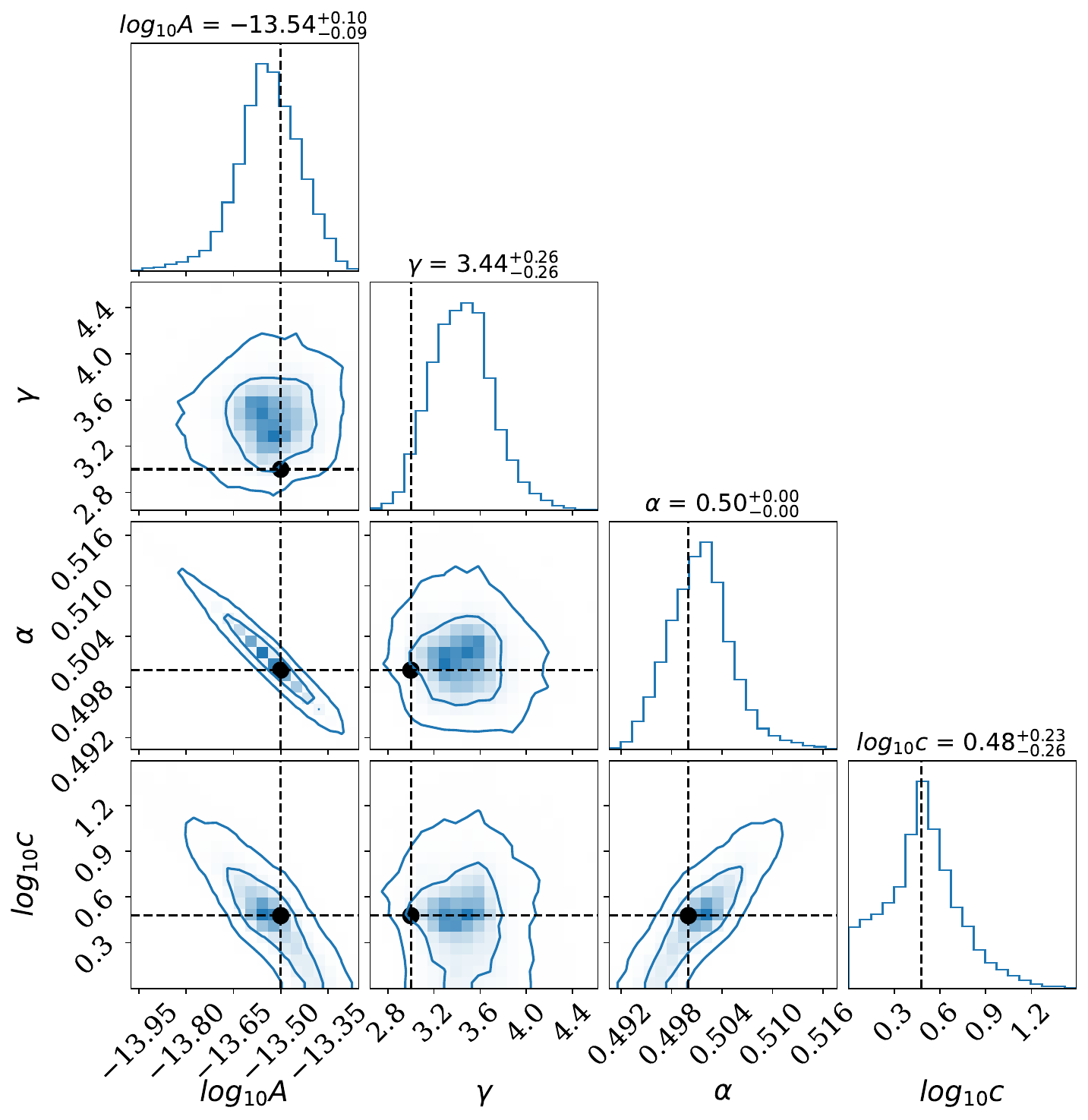}
    \caption{Corner plot of the recovered parameters for the non-Gaussian model in single pulsar noise analysis. The Fourier coefficients are also sampled but not shown here for simplicity. The dashed line show the injected parameter values. The shaded areas show the 1 and 2 $\sigma$ credible regions.}
    \label{fig:spa_corner}
\end{figure}

The result of one simulation is shown in \autoref{fig:spa_corner} to verify that we indeed recover the injected values of  the parameters. We only show the hyperparameters characterizing the noise, while numerically marginalizing over all Fourier coefficients by sampling them. We have used 50 frequencies, for a total of 100 Fourier coefficients that we fit for. This approach is straightforward as it requires no analytical marginalization. However, the sampling is affected by the high dimensionality of the parameter space and many samples are required to converge to the proper distribution. Still, in the case of GMMs, the non-marginalized (method I) and marginalized (method II) likelihoods presented in \autoref{sec:ng_likelihood} should respond equally to non-Gaussianities present in the data. This can be seen in \autoref{fig:crn_per_kurt_bf} where we plot the Bayes factor $\mathcal{B}^{NG}_G$ comparing the non-Gaussian and Gaussian models as a function of the excess kurtosis $\Delta \bar{M}_4$. As $\Delta \bar{M}_4$ increases, the Bayes factor increases accordingly. Note that for $\Delta \bar{M}_4 = 0$ (i.e. Gaussian noise is injected), we have $\mathcal{B}^{NG}_G \approx 1$ meaning that the non-Gaussian model is not preferred over the Gaussian model.

\subsection{Common red noise}

We apply the per-frequency marginalized non-Gaussian likelihood method to infer the statistics of the injected common noise in the 100 pulsars simulated dataset. Our goal is to test that the Bayes factor correctly favors the injected non-Gaussianities over a Gaussian model while recovering the injected parameters. We simulate an uncorrelated common red noise where each frequency bin $n$ follows a 2 Gaussian GMM $\vec{X}_n, \vec{Y}_n \sim p(\alpha, c, S(A, \gamma, f_n)\Delta f)$. The noise spectrum has characteristics similar to what is expected for a GWB produced by a population of SMBHBs in circular orbit with characteristic strain $h_c = A (f/f_{yr})^\beta$ with $\beta=-2/3$ and $A=10^{-15}$. The conversion from $h_c$ to the PSD of the induced timing residuals is given by $S(f)=h^2_c/(12\pi^2 f^3)$ \cite{epta_wm5}, hence $\gamma=3 - 2\beta=13/3$.

We fix $\alpha=0.5$, $\gamma=13/3$ and vary $c$ between $1$ and $15$ to control the excess moments of the distribution. We sample the posterior distribution of  \autoref{eq:posterior_prob} constructed from the likelihood in \autoref{eq:ng_likelihood} using MCMC techniques. We assess the significance of non-Gaussianities by calculating the Bayes factor for the non-Gaussian model versus the Gaussian model $\mathcal{B}^{NG}_G$. The latter is estimated using hypermodeling \cite{Carlin_Chib, hee_hypermodel} (product-space sampling) or alternatively, with the Savage-Dickey density  since the Gaussian model is obtained by fixing $\alpha=0$ \cite{Arzoumanian_2018, Hazboun_2020}. We perform per-frequency free spectrum fits, modeling non-Gaussian features characterized by parameters $\Phi$, $\alpha$ and $c$. The prior ranges of the parameters are listed in \autoref{tab:prior_ranges}. We produced a PP-plot to assess the good calibration of the simulations (see \autoref{fig:pp_plot} in Appendix ~\ref{app:pp}).

In \autoref{fig:crn_per_kurt_bf} we show $\mathcal{B}^{NG}_G$ as a function of the injected excess kurtosis $\Delta \bar{M}_4$ estimated using \autoref{eq:excess_moments}. We see a clear increase of the Bayes factor for growing $\Delta \bar{M}_4$, indicating that non-Gaussianities can be detected in Bayesian model selection.

\begin{figure}[h!]
\centering
    \includegraphics[width=0.45\textwidth]{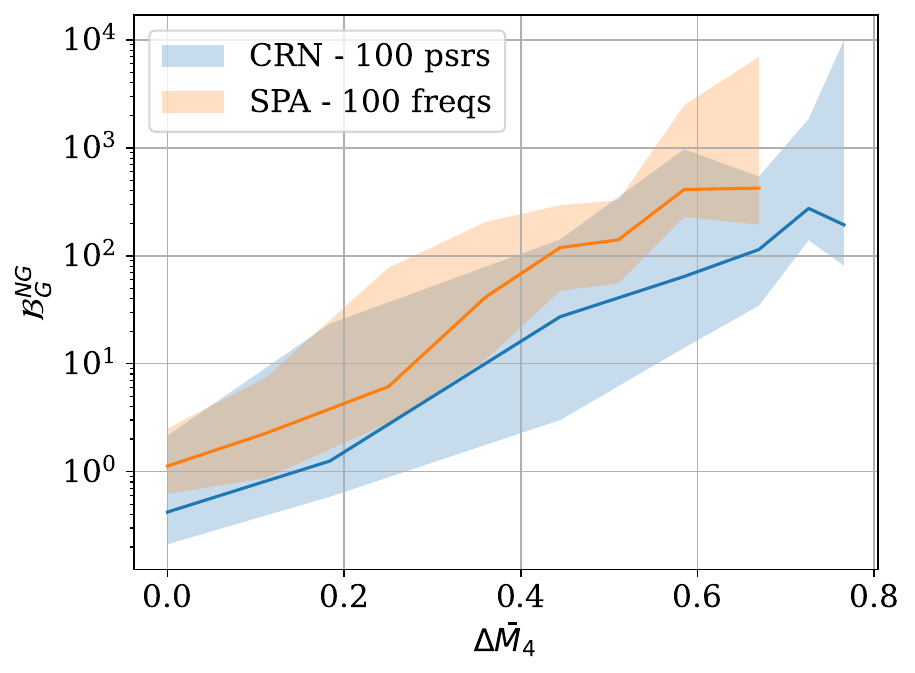}
    \caption{Bayes factor $\mathcal{B}^{NG}_G$ between the non-Gaussian and Gaussian models as a function of the injected relative excess kurtosis $\Delta \bar{M}_4$ calculated using \autoref{eq:excess_moments} for many simulations. The solid line is the median and the shaded region is bounded by the 16th and 84th percentiles. We show the two cases : (i) SPA (Single Pulsar Analysis) with 100 frequencies powerlaw red noise, (ii) CRN (Common red noise) with 100 pulsars.}
    \label{fig:crn_per_kurt_bf}
\end{figure}

\begin{figure}[h!]
\centering
    \includegraphics[width=0.45\textwidth]{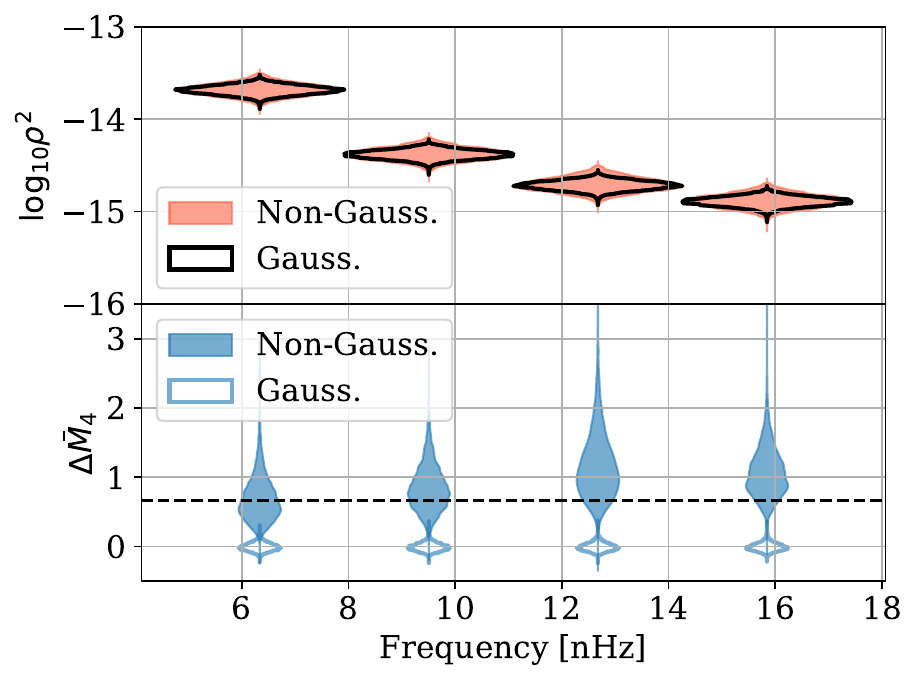}
    \caption{Recovered free spectrum and excess kurtosis for the Gaussian model and non-Gaussian model at the 2, 3, 4, 5 harmonics of 10yrs for the simulated data containing a non-Gaussian common red noise following a powerlaw spectrum with $\log_{10} A=-15$, $\gamma=13/3$, $\alpha=0.5$ and $c=10$. \textit{(Top panel)} Violin plot of the estimated free spectrum, the two models are very consistent. \textit{(Bottom panel)} Violin plot of the excess kurtosis posterior distribution giving zero for the Gaussian model, the dashed line shows the injected excess kurtosis obtained with \autoref{eq:excess_moments}, that is the same for all frequencies.}
    \label{fig:fs_crn}
\end{figure}

\begin{figure}[h!]
\centering
    \includegraphics[width=0.45\textwidth]{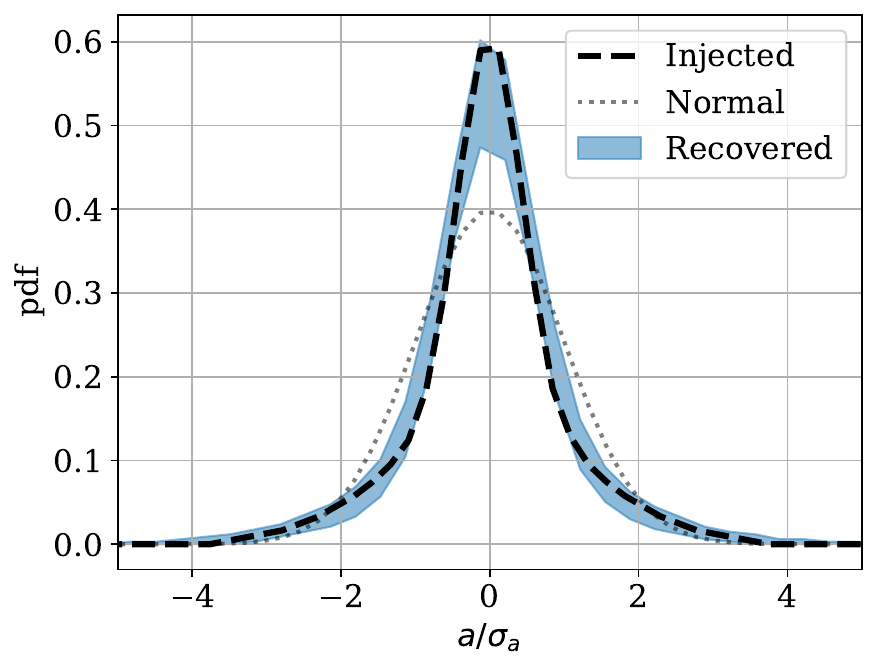}
    \caption{Comparison between the injected and recovered probability distribution of the Fourier coefficients $a$, divided by the standard deviation $\sigma_a$, showing the deviation from Gaussianity. The blue shaded region corresponding to the binned 1-$\sigma$ error obtained with 1000 reconstructions of $p(a_n, \Phi_n, \alpha,c|\delta t)$ using samples from $p(\Phi_n, \alpha,c|\delta t)$ according to \autoref{eq:posterior_ng}. The dotted line shows a standard normal distribution for comparison.}
    \label{fig:hist_ng}
\end{figure}

The kurtosis increase can be seen in \autoref{fig:fs_crn} where the 2nd and 4th moment posterior distributions given by \autoref{eq:qth_order_fs} are showed for the Gaussian and non-Gaussian models. We see that the second moment is the same for both, meaning that the data informs the correct PSD under the hypothesis of Gaussian noise. However, the latter is unable to detect higher order statistics that could contain interesting information about the signal. The relative excess kurtosis in \autoref{fig:fs_crn} is calculated using \autoref{eq:excess_moments} and \autoref{eq:qth_order_fs} as $\Delta \bar{M}_4 = [\rho^4 - 3(\rho^2)^2]/3(\rho^2)^2$. In \autoref{fig:hist_ng}, we show the reconstruction of the Fourier coefficients statistic using \autoref{eq:posterior_ng} by drawing samples from the posterior distribution of parameters $\Phi_n$, $\alpha$ and $c$. We correctly recover the injected probability distribution within 1-$\sigma$ errors. We see that it has thinner body and heavier tails than a Gaussian, matching the expected shape for excess kurtosis.

\subsection{Background plus few sources}

An astrophysically interesting scenario where non-Gaussianities can be observed is when the signal in the data consists of a  Gaussian background noise plus a few loud individual sources at some specific frequency \cite{racine_ng}. This is, in fact, a reasonable proxy for what is expected from a realistic population of SMBHBs, since some binaries will be closer to us or more massive, appearing in the data as relatively loud CGWs \cite{2009MNRAS.394.2255S,babak_cgw, epta_cgw, ng_cgw, ipta_cgw}. If these binaries are identified and subtracted, the background population should be isotropic enough to produce a Gaussian noise, as expected by the law of large numbers \cite{racine_ng}. In the following, we investigate the non-Gaussian statistics emerging from the superposition of a Gaussian GWB with a few CGWs. In order to apply Method 2 of \autoref{sec:ng_likelihood}, we study the case where the CGWs fall within the same frequency bin. In PTA, an individual, circular SMBHB produces timing residuals given by the difference between the signal seen on Earth and at the pulsar 

\begin{equation}
    s_a(t) = h \sum_{A=+,\times} F_A(\hat{p}_a, \hat{k}) [s_A(t) - s_A(t-\tau_a)]\,,
\label{eq:cgw_wf}
\end{equation}
where $h$ is the GW amplitude, $F_A$ the antenna pattern function, $\hat{p}_a$ the unit vector pointing at the position of pulsar $a$, $\hat{k}$ the direction of propagation of the wave, $s_A(t)$ the circular binary waveform for each polarization modes $+$ and $\times$ that is a function of the frequency $f_{gw}$ of the wave, the polarization angle and the inclination of the binary, and $\tau_a$ the relative time delay between Earth time and pulsar time (for more details about this type of signal and its full mathematical expression, see \cite{epta_cgw, ng_cgw, babak_cgw, ipta_cgw}).

We inject the same GWB as in the previous section with characteristic strain $h_c=10^{-15} \left( {f}/{f_{\rm yr}} \right)^{-2/3}$. We inject in the data one, two or three CGWs with a frequency of $3/(10 {\rm yr})$ according to \autoref{eq:cgw_wf} and check for Gaussianity of the common noise at this frequency. The SNR of the sources with respect to the GWB is calculated in time-domain as

\begin{equation}
    SNR^2=\sum_{a=1}^{N_p} s_a^\top C^{-1} s_a,
\end{equation}
where $C(t - t')=\sigma^2 \mathcal{I}_{N_t} + \sum_n S(f_n) \Delta f\cos[2\pi f_n(t - t')]$.

\begin{figure}[h!]
\centering
    \includegraphics[width=0.45\textwidth]{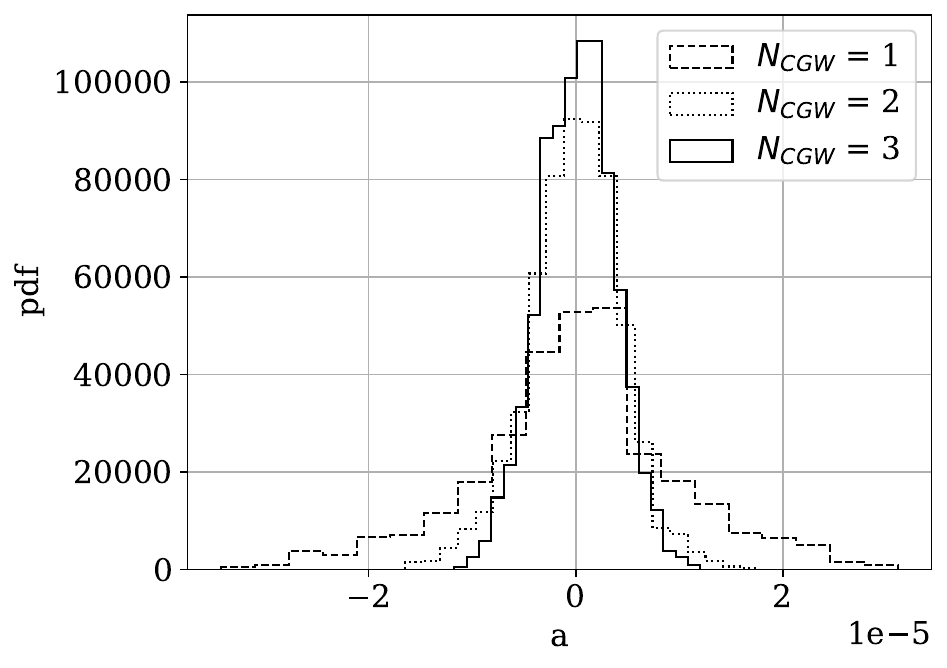}
    \caption{Histogram of the FFT coefficients calculated for each pulsar timing residuals at frequency 3/(10yr) where a common red noise and 3 CGW sources were injected in sequence. We gradually added the three CGWs with amplitudes of $\log_{10}h=-14.5$, $-14.65$ and $-15$ corresponding to SNRs of, respectively, 37, 17, 7. The value $N_{CGW}$ marks the total number of CGWs injected in the data. By adding more CGWs the tails of the distribution get progressively damped.}
    \label{fig:fourier_cw}
\end{figure}

To illustrate the effect of single sources in the statistic of the Fourier coefficients, in \autoref{fig:fourier_cw}, we show the distribution of the Fourier coefficients at frequency $3/(10{\rm yr})$ obtained with the Fast Fourier Transform (FFT) of simulated pulsar timing data where a GWB and some CGW sources are injected. Each pulsar FFT gives two values (real and imaginary part at the considered frequency) and we plot a histogram of all combined coefficients across pulsars. We see that a single CGW produces a heavy-tailed distribution, that becomes progressively more Gaussian when adding few more CGWs.

\begin{figure}
\centering
    \includegraphics[width=0.45\textwidth]{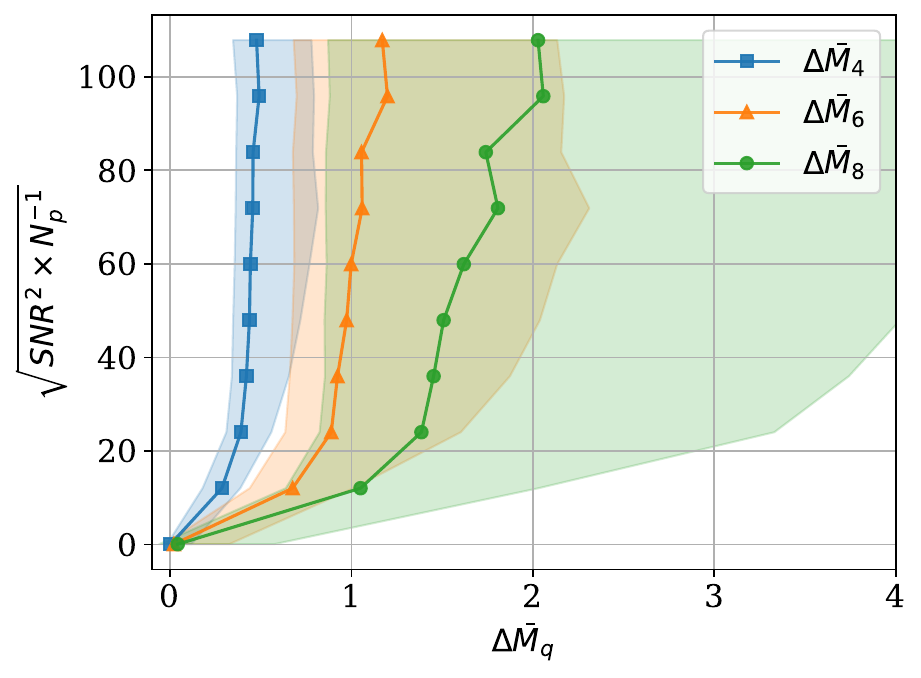}
    \caption{Excess moments as a function of the averaged SNR of the source for $N_p$ pulsars with $10$ years of observation, $100$ ns noise level and a GWB signal with $h_c = 10^{-15} (f/f_{\rm yr})^{-2/3}$. Since $f_{\rm gw} = 3/(10\textrm{yr})$, the source produces 3 cycles in the data. This plot shows the median excess moments estimated from a 100 realization of the CGW parameters. The shaded colored areas show the 16th and 84th percentiles of the realizations. Only the amplitude of the CGW was fixed to control the SNR.}
    \label{fig:snr_moments}
\end{figure}

\begin{figure}
\centering
    \includegraphics[width=0.45\textwidth]{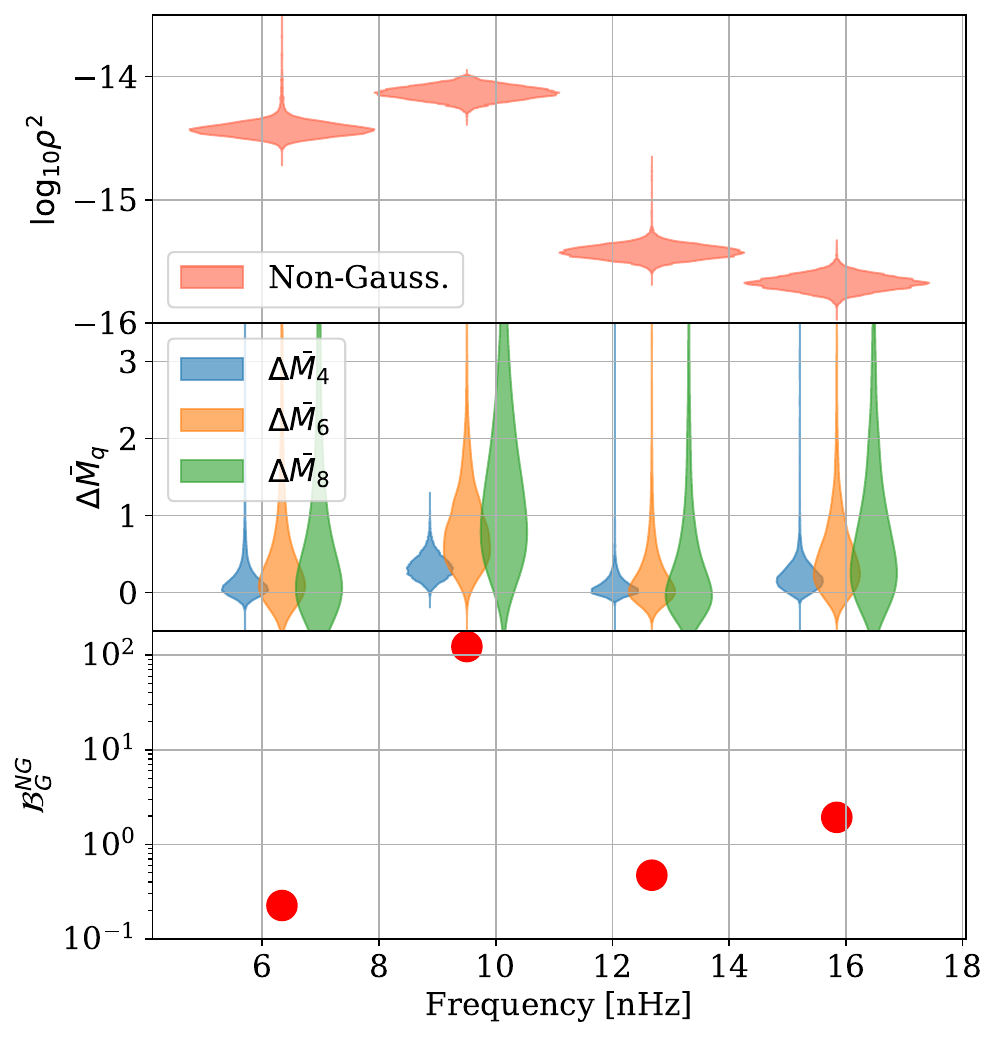}
    \caption{Recovered free spectrum, excess moments and Bayes factor for the non-Gaussian model at the 2nd, 3rd, 4th, 5th harmonics of $f=1/10$yr for the simulated data containing a CGW at frequency 3/(10yr). \textit{(Top)} violin plot of the estimated free spectrum, the 3/(10yr) shows the excess power due to the presence of the CGW \cite{Ferranti_2025}. \textit{(Middle)} violin plot of the excess moment posterior distributions, the uncertainties are large but we see that the frequency 3/(10yr) is well constrained to a higher value than other bins, showing that the model catches the excess of power in higher moments introduced by the CGW. \textit{(Bottom)} the Bayes factor $\mathcal{B}^{NG}_G$ at each frequency shows that the model favors non-Gaussianity where the CGW is present.}
    \label{fig:fs_crn_cgw}
\end{figure}

The fact that adding a single source introduces non-Gaussianities is illustrated in \autoref{fig:snr_moments}, where we show the added higher order moments as a function of the SNR of that source, directly calculated from the histograms of the FFT coefficients for different CGW parameter realizations as shown in \autoref{fig:fourier_cw}. In particular, we show the relative excess moments with respect to a Gaussian as defined in \autoref{eq:excess_moments}. It is clear that the deviations from Gaussianity grow with the SNR. This has to do with the pulsar response to GWs $F^{+,\times}$ introduced in \autoref{eq:cgw_wf}. When one CGW is present in the data, it is going to appear very bright only in some pulsars that respond well to it, usually the pulsars close to the sky location of the source \cite{boyle_and_pen}. This is what causes the heavier tails in the distribution of the Fourier coefficients in \autoref{fig:fourier_cw} and puts more power in the higher order moments in \autoref{fig:snr_moments}. As more CGW sources are present and isotropically distributed in the sky, all the pulsars will equally respond to a signal. Because the signals they see have random phases, for a large number of CGWs, the central limit theorem tells us that the Fourier coefficients will be asymptotically Gaussian distributed and spatially correlated following the HD correlation pattern \cite{bruce_variance}. Then, there is an intricate relation between the anisotropy of an astrophysical GWB and its non-Gaussianity.

We test our model on the scenario where only one CGW is injected with an amplitude $\log_{10} h = -14.25$ giving $\rm SNR=50$ \footnote{The SNR is high because it scales as $\sqrt{N_p}$ and we simulate $N_p=100$ equally good pulsars. Even though a high SNR should yield more informative posterior distributions, it does not systematically mean that the injected excess moments $\Delta \bar{M}_q$ are very large, since the latter depend on the statistics of Fourier coefficients in individual pulsars. To that extent, $\Delta \bar{M}_q$ should scale as the average of the SNRs calculated for individual pulsars.}.
In \autoref{fig:fs_crn_cgw} we show the posterior distribution for higher order moments calculated with \autoref{eq:qth_order_fs} and \autoref{eq:excess_moments}. We notice that the second bin of the free spectrum in the top panel shows and excess of power due to the presence of the CGW \cite{Ferranti_2025}. The middle panel shows that the posterior distributions for higher order moments are consistent with zero, except where the CGW is present and the Bayes factor favors the non-Gaussian model over the Gaussian model. Still, the error bars are quite large, and the model does not seem to fully catch the characteristics of the underlying distribution. This is because a mixture of two Gaussians is a limited model that will saturate in the amount of information it can capture. Finding the optimal number of Gaussians in the GMM will be essential for future developments.

\begin{figure*}[t!]
  \centering
  \begin{minipage}[t]{0.49\textwidth}
    \centering
    \includegraphics[width=\linewidth]{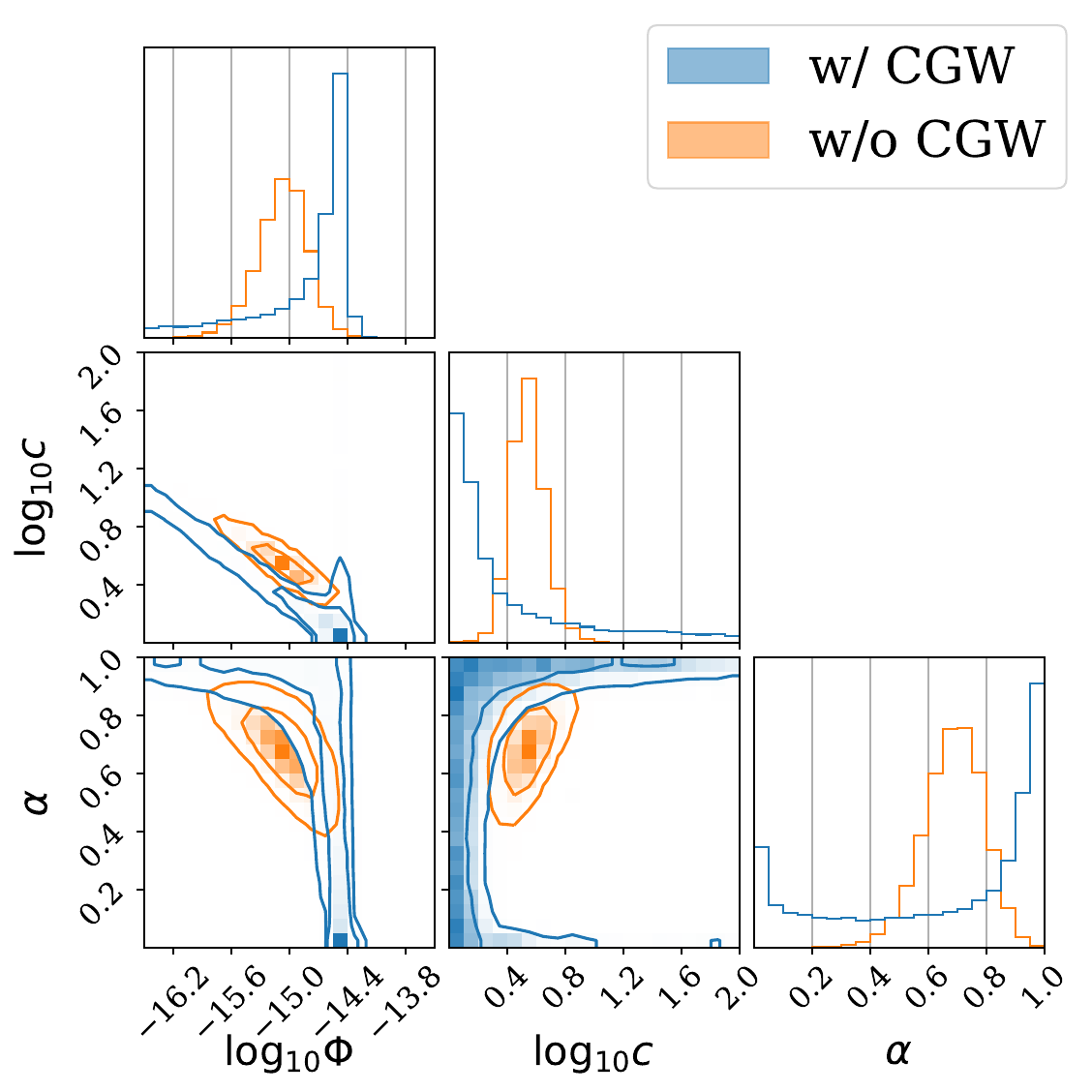}
  \end{minipage}
  \hfill
  \begin{minipage}[t]{0.49\textwidth}
    \centering
    \includegraphics[width=\linewidth]{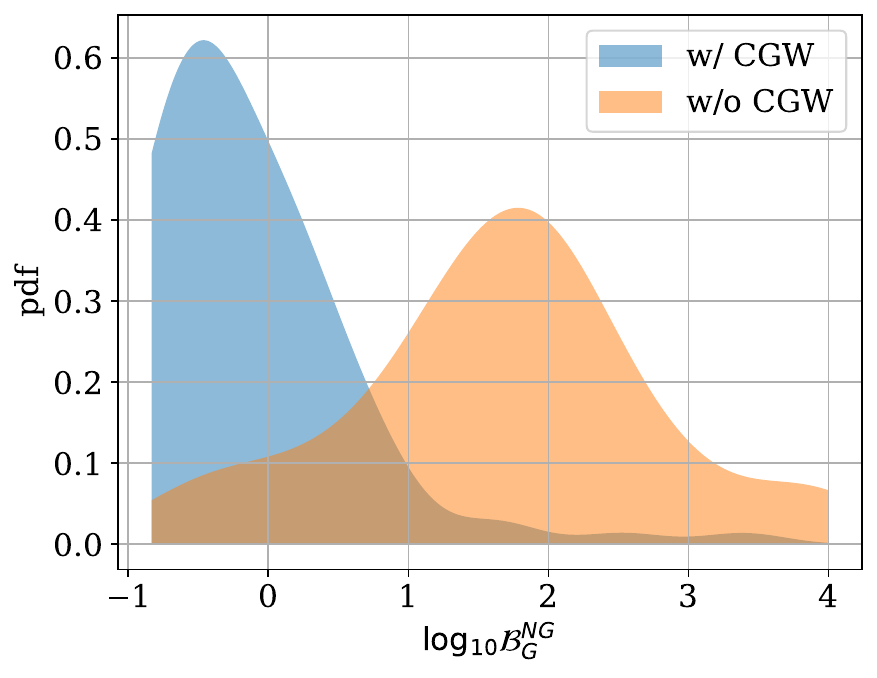}
  \end{minipage}
  \caption{(\textit{left}) Corner plot of the non-Gaussianity parameters obtained at frequency 3/(10yr) for a model with CGW as a deterministic signal (blue histogram) and without the CGW (orange histogram). The shaded areas show the 1 and 2 $\sigma$ credible regions. While the orange histogram is well constrained around $\log_{10} c \approx 0.5$ and $\alpha \approx 0.7$, the blue one is bimodal and consistent with $\alpha=0$ or $\alpha=1$ and $\log_{10} c = 0$, showing that the model favors only one Gaussian in the mixture (\textit{right}) Distribution of the recovered Bayes factors $\mathcal{B}^{NG}_G$ for 100 simulated datasets containing Gaussian noise with a single source at frequency 3/(10yr). The significance of non-Gaussianities in the data is high when the CGW is not accounted for in the model (orange histogram), but the inclusion of the CGW as a deterministic signal in the model destroys this significance (blue histogram).}
  \label{fig:w_wo_cgw}
\end{figure*}

In \autoref{fig:w_wo_cgw} we show the recovered posterior distribution for the non-Gaussianity parameters when we include or not the CGW as a deterministic signal in the model used in the recovery. We choose a realization of the data where the CGW introduced significant non-Gaussianities, again modeled as the mixture of 2 central Gaussians controlled with parameters $\alpha$ and $\log_{10} c$. The introduction of the deterministic CGW signal in the model removes the non-Gaussian features and destroy their significance. For the model without CGW, we find posterior median values of $\alpha=0.7$ and $\log_{10} c=0.5$ while the inclusion of the deterministic CGW gives posteriors consistent with $\alpha=0$, or $\alpha=1$ and $\log_{10} c=0$, corresponding to the solutions where only one Gaussian of the mixture is active (i.e. Gaussian behavior is favored). Moreover, the recovered Bayes factor drops from $\mathcal{B}^{NG}_G\approx100$ to $\mathcal{B}^{NG}_G\approx 0.5$, showing that the significance is greatly impacted by the subtraction of the CGW waveform in the timing residuals (see \autoref{fig:w_wo_cgw}). This fact could be used to assess the significance of the presence of individual sources in the data by measuring how including the correct CGW signal in the model correctly yields a Gaussian likelihood.

\begin{figure*}[t!]
  \centering
  \begin{minipage}[t]{0.49\textwidth}
    \centering
    \includegraphics[width=\linewidth]{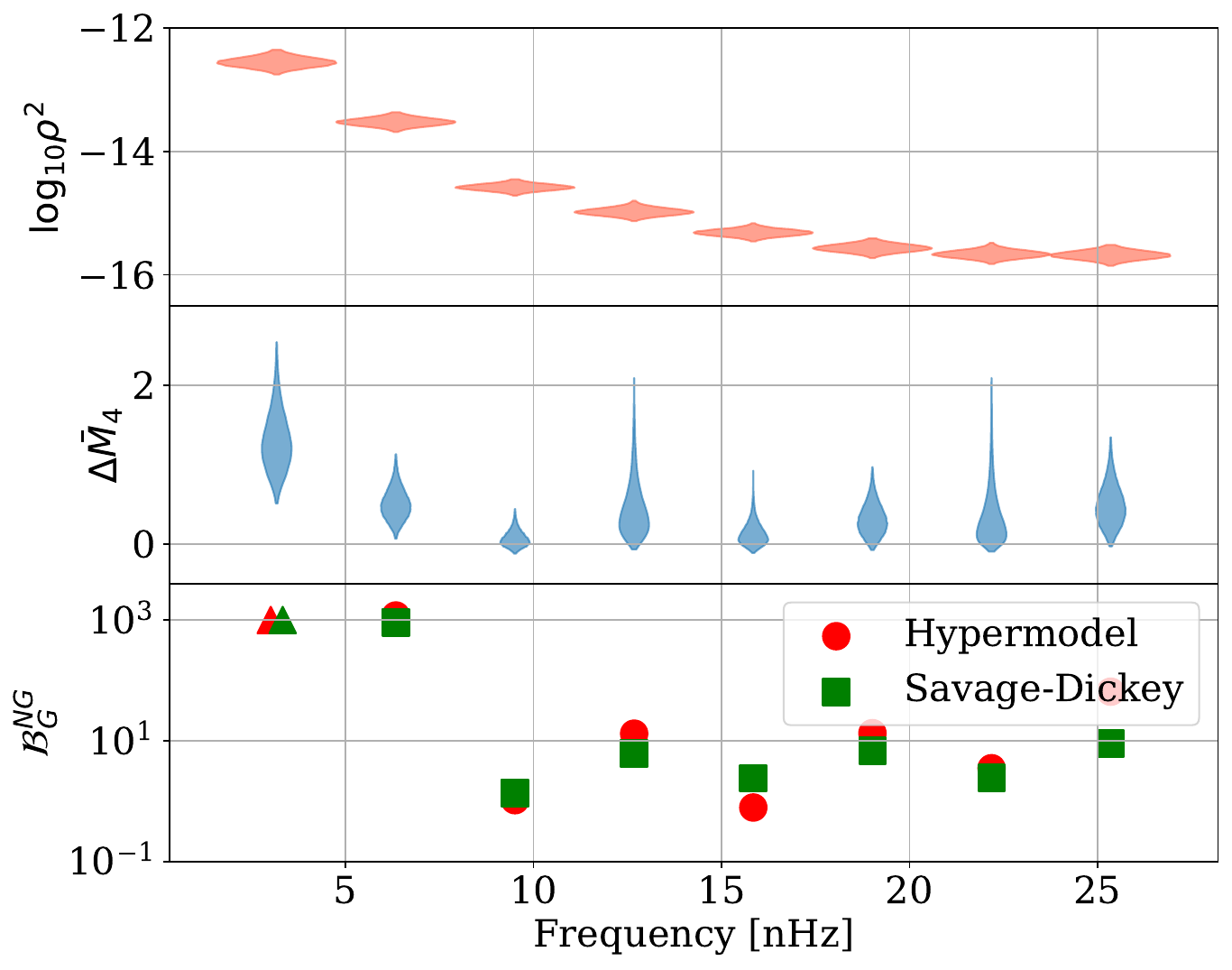}
  \end{minipage}
  \hfill
  \begin{minipage}[t]{0.49\textwidth}
    \centering
    \includegraphics[width=\linewidth]{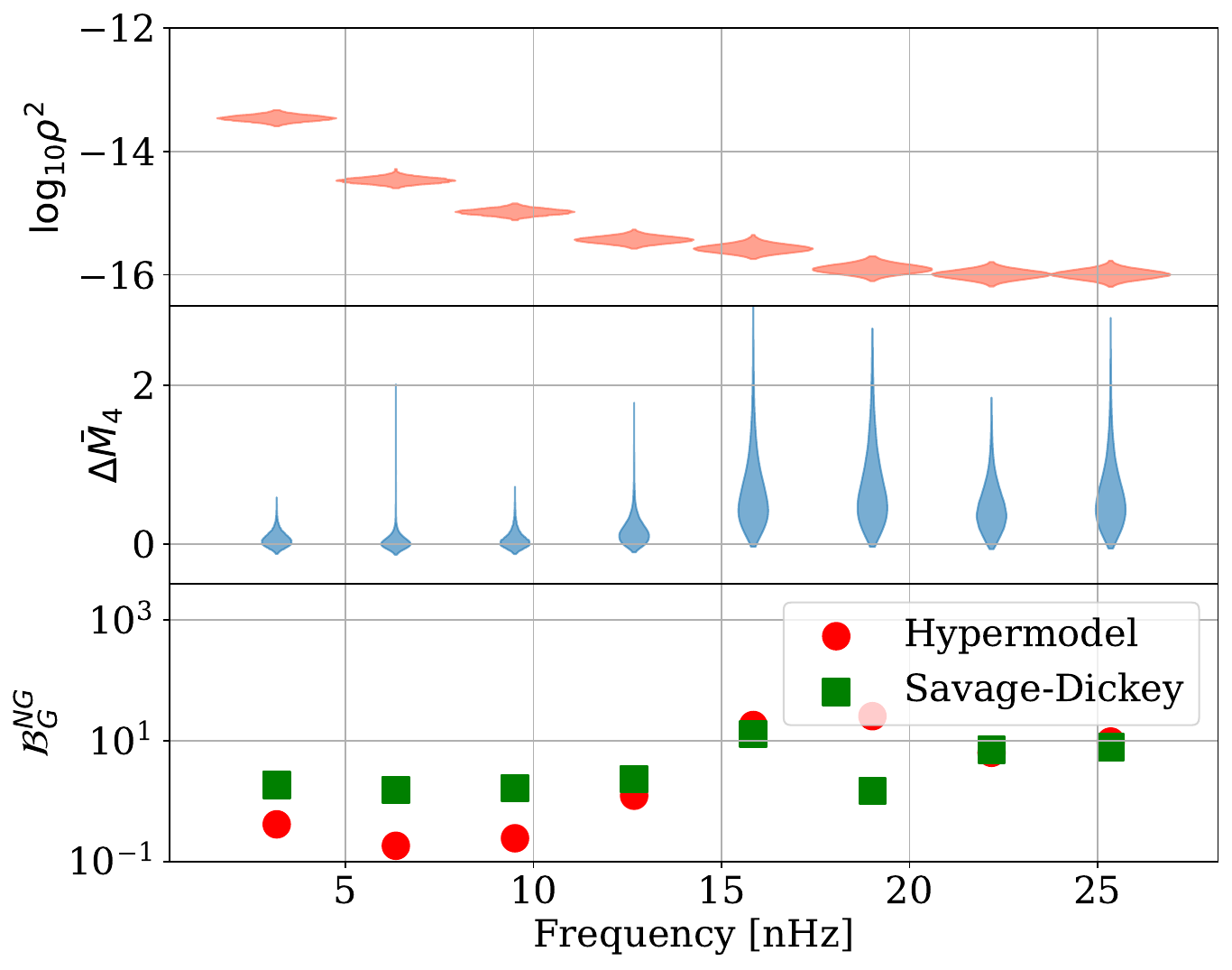}
  \end{minipage}
  \caption{Posterior distributions for the free spectrum $\rho^2$, excess kurtosis $\Delta \bar{M}_4$ and Bayes factor $\mathcal{B}^{NG}_G$ as a function of frequency inferred from a simulated PTA with a realistic GWB signal (\textit{left}) with a resolved CGW signal injected at 5 nHz (\textit{right}) without any CGW signal injected. The Bayes factor obtained with the two methods are roughly consistent. The triangles for $\mathcal{B}^{NG}_G$ in the first bin of the left panel indicate that the recovered $\mathcal{B}^{NG}_G$ is far greater than $10^3$ and difficult to precisely estimate numerically. Both method yielded very high values ranging between $10^6$ and $10^{12}$.}
  \label{fig:real_gwb}
\end{figure*}

\subsection{Realistic astrophysical background}

Finally, we generate a realistic astrophysical GWB by estimating the individual CGW emissions of a population of SMBHBs. The injection pipeline uses a simulated catalog of SMBHBs providing their chirp mass, frequency, distance, inclination, polarization and sky location. Their contributions are added one by one using \autoref{eq:cgw_wf} in a 100 pulsars array with 10 years of observation and only white noise at a level of $10^{-7}$s. The catalogs that are used are the same as in \cite{Ferranti_2025} to produce the simulations referred to as \textbf{dataset injGWB03} and \textbf{dataset injGWB03+injCGW5}. Both contain the same realization of a GWB signal but the second has a CGW emitting at 5nHz, resolved on top of the GWB (for more details, see section 2.1.2 of \cite{Ferranti_2025}). The datasets are analyzed using the non-Gaussian model constructed from a GMM of 2 Gaussians controlled by the parameters $\alpha$ and $c$. We estimate Bayes factors with hypermodeling \cite{hee_hypermodel} or, using the fact that one can turn off non-Gaussianities by setting $\alpha=0$, with the Savage-Dickey density ratio as \cite{Arzoumanian_2018}

\begin{equation}
    \mathcal{B}^{NG}_G= \frac{\Pi_\alpha(\alpha = 0)}{p_\alpha (\alpha=0)},
\end{equation}
where $\Pi_\alpha(\alpha = 0)=1$ is the prior probability distribution for parameter $\alpha$ evaluated at 0, and $p_\alpha (\alpha=0)$ is the posterior probability distribution for $\alpha$ evaluated at 0.

In \autoref{fig:real_gwb} we show the recovered free spectrum, excess kurtosis and Bayes factor $\mathcal{B}^{NG}_G$ for the first 8 harmonics of $1/(10 \rm yrs)$. We display only the excess kurtosis because the higher order moments have very broad posteriors that are not very informative. The free spectra between the two simulations differ from the injection of the CGW at 5 nHz. Most of the injected power by the CGW is present in the first two frequency bins and appears to slightly leak at higher frequencies\footnote{This is due to the finite time of observation that transforms sinusoids in time domain into sinc functions in frequency domain, causing leakage between neighboring frequencies \cite{hazboun_2019}.}. The most striking difference is when looking at the $\Delta \bar{M}_4$ posteriors where the 5 nHz CGW injects a significant excess kurtosis in the first two frequencies, and the Bayes factor favoring the presence of non-Gaussianities is high ($\mathcal{B}^{NG}_G > 10^3$), as expected from what is discussed in the previous section. The simulation without CGW does not show significant signs of non-Gaussianities, especially at low frequencies where $\Delta \bar{M}_4$ is well constrained around zero and $\mathcal{B}^{NG}_G < 1$. This can be explained by the higher number of contributing GW sources at low frequencies, producing a more isotropic and Gaussian GWB signal. Both $\Delta \bar{M}_4$ and $\mathcal{B}^{NG}_G$ seem to slightly increase for higher frequencies, as the number of contributing GW sources decreases, which might enhance the non-Gaussianities in the spectrum. The detection of mild non-Gaussian behavior in PTA might require a larger number of pulsars to increase our sensitivity to the statistic of the spectrum.

\section{Conclusion}

In this work, we introduced a method to model non-Gaussian features in PTA data by using GMMs as priors for the Fourier coefficients of stochastic processes. This approach enables analytical marginalization of the likelihood, making it computationally efficient and directly compatible with existing PTA analysis pipelines. We validated the method through simulations involving both single pulsar noise and common red noise processes, demonstrating that the model can robustly identify departures from Gaussianity using Bayesian model selection. The Bayes factor increases significantly in the presence of excess kurtosis, confirming the sensitivity of the approach to higher-order statistical structure. Furthermore, we applied the method to a more astrophysically motivated scenario involving a superposition of a Gaussian GWB and a few CGW sources. In this case, we showed that non-Gaussian features become detectable when a small number of loud sources dominate the signal, and that these features are suppressed once the deterministic CGWs are correctly modeled and subtracted. When a realistic GWB is generated from the superposition of the individual CGW emissions of a simulated population of SMBHBs, the non-Gaussianities are harder to detect and might require very large arrays of pulsars to increase their detectability. This is illustrated in Appendix ~\ref{app:bf_npsrs} where we show the Bayes factor $\mathcal{B}^{NG}_G$ for different sizes of array and varying excess kurtosis.

One of the key outcomes of this study is the demonstration that posterior estimates of higher-order central moments, particularly the fourth moment (kurtosis), can provide complementary information beyond the PSD, which is typically inferred under the assumption of Gaussianity. By constructing posterior distributions for the higher-order moments from the Fourier coefficients, the model can capture statistical features such as heavier tails or asymmetries in the signal that are not accounted for in standard analyses. This opens up the possibility of using spectral refitting techniques that explicitly include these moments as informative observables \cite{Lamb_2023, Lamb_2024}. In particular, the excess moments may carry physical information about the population of GW sources, for example, indicating the presence of a few dominant binaries within an otherwise Gaussian background. Incorporating this information could help refine the characterization of the astrophysical origin of the signal and improve constraints on the source population models, especially in the intermediate regime where the number of contributing sources is not large enough for the central limit theorem to fully apply.

Although we focused here on mixtures of two centered Gaussian distributions and considered spatially uncorrelated physical processes for simplicity and computational tractability, the framework can be readily extended. Future work could explore non-symmetric and multi-modal mixtures to capture a broader class of non-Gaussian behaviors, including skewness. Additionally, methods such as reversible-jump MCMC could be employed to dynamically infer the number and properties of the mixture components. The need to implement a version of this analysis accounting for the interfrequency spectral correlations due to the finite time of observation will be essential for the future, as highlighted in \cite{crisostomi2025}. From a modeling perspective, the inclusion of higher-order inter-pulsar correlation structures (4-point correlator) through Gram-Charlier or Edgeworth expansions has not yet been implemented due to its increased computational cost. Finally, the integration of these techniques into full Bayesian PTA analysis pipelines may provide a novel diagnostic to disentangle the statistical properties of astrophysical signals, potentially offering a route to discriminate between different GW source populations and to detect outliers or individual resolvable binaries within a stochastic background.

\begin{acknowledgments}
We thank Irene Ferranti for providing the realistic population catalogs. acknowledge support from the European Union’s H2020 ERC Advanced Grant ``PINGU'' (Grant No. 101142079) and from the PRIN 2022 grant "GRAPE" (CUP H53D2300091
0006).
\end{acknowledgments}

\begin{widetext}

\appendix

\section{Moments of a Gaussian mixture}
\label{app:moments}

A mixture of 2 Gaussians can be expressed using \autoref{eq:gaussian_mixture} as

\begin{equation}
    p(x) = (1 - \alpha) \varphi(x, 0, \sqrt{\Phi}) + \alpha \varphi(x, \mu_0, \sqrt{c\Phi}),
\end{equation}
with mean $\int x p(x)dx = \alpha \mu_0$. Its central moments are given by

\begin{equation}
\begin{aligned}
    M_n & = \int (x - \alpha \mu_0)^n p(x) dx\\
    & = (1 - \alpha) \int (x - \alpha \mu_0)^n \varphi(x, 0, \sqrt{\Phi}) dx+ \alpha \int (x - \alpha \mu_0)^n \varphi(x, \mu_0, \sqrt{c\Phi})dx\\
    & = (1 - \alpha) \int \tilde{x}^n \varphi(\tilde{x}, -\alpha \mu_0, \sqrt{\Phi}) d\tilde{x}+ \alpha \int \tilde{x}^n \varphi(\tilde{x}, (1 - \alpha)\mu_0, \sqrt{c\Phi})d\tilde{x}\\
    & = (1 - \alpha) \mathcal{M}_n(-\alpha \mu_0, \sqrt{\Phi}) + \alpha \mathcal{M}_n ((1-\alpha) \mu_0, \sqrt{c \Phi}),
\end{aligned}
\end{equation}
with $\tilde{x} = x - \alpha \mu_0$ and $\mathcal{M}_n(m, \sigma)$ the raw (non-central) moments of a normal distribution with mean $m$ and standard deviation $\sigma$ that are well-known and can be found in the literature. The first four $M_n$ are given in \autoref{tab:moments}.

\section{Marginalized Gaussian likelihood}
\label{app:margin_likelihood}

Consider a dataset $\delta t$ containing uncorrelated Gaussian white noise characterized by a diagonal noise matrix $N$. We want to model a signal $s=Fa$ decomposed on a basis $F$ (typically a discrete Fourier basis), characterized by a vector of coefficients $a$. The coefficients $a$ follow a Gaussian distribution with covariance $\Phi$. The likelihood for this model can be written as

\begin{equation}
    \mathcal{L}(\delta t|a,\Phi, \mu)=\frac{\exp \left \{ -\frac{1}{2} \left [\delta t - F a \right ] ^\top N^{-1} \left[\delta t - F a \right ] \right \}}{|2\pi N |^{1/2}} \times \frac{\exp \left \{ -\frac{1}{2} [a - \mu] ^\top \Phi^{-1} [a - \mu ] \right \} }{|2\pi \Phi |^{1/2}}
\end{equation}
where on the left we have the Gaussian likelihood for uncorrelated white noise and on the right, the prior probability distribution for coefficients $a$. Developing this expression, we get

\begin{equation}
    \mathcal{L}(\delta t|a, \Phi, \mu)=\frac{\exp \left \{ -\frac{1}{2} \delta t^\top N^{-1} \delta t -\frac{1}{2} a^\top \left[ F^\top N^{-1} F + \Phi^{-1} \right] a + a^\top \left[F^\top N^{-1} \delta t + \Phi^{-1} \mu \right] \right \} }{|2\pi N |^{1/2}|2\pi \Phi |^{1/2}} \times \exp \left \{-\frac{1}{2} \mu^\top \Phi^{-1} \mu \right \}.
\end{equation}

Defining $\Sigma = F^\top N^{-1} F + \Phi^{-1} $ and completing the squares in the numerator we get

\begin{equation}
\begin{aligned}
    & -\frac{1}{2} a^\top \Sigma a + a^\top \left[F^\top N^{-1} \delta t + \Phi^{-1} \mu \right]\\
    = & -\frac{1}{2} \left[ a - \Sigma^{-1} F^\top N^{-1} \delta t - \Sigma^{-1} \Phi^{-1} \mu\right]^\top \Sigma \left[ a - \Sigma^{-1} F^\top N^{-1} \delta t - \Sigma^{-1} \Phi^{-1} \mu \right] \\
    & + \frac{1}{2} \left[\mu^\top \Phi^{-1} \Sigma^{-1} + \delta t ^\top N^{-1} F \Sigma ^{-1} \right] \Sigma \left[\Sigma^{-1} F^\top N^{-1} \delta t + \Sigma^{-1} \Phi^{-1} \mu \right]
\end{aligned}
\end{equation}

Defining $\hat{a} = \Sigma^{-1} F^\top N^{-1} \delta t + \Sigma^{-1} \Phi^{-1} \mu$ and re-arranging the terms we get

\begin{equation}
\begin{aligned}
    \mathcal{L}(\delta t | a, \Phi, \mu) = & \frac{\exp \left \{ - \frac{1}{2} \delta t^\top \left[N^{-1} - N^{-1} F \Sigma^{-1} F^\top N^{-1} \right] \delta t \right \}}{|2\pi N|^{1/2} |2\pi \Phi|^{1/2}} \times \exp \left \{ -\frac{1}{2} \left[a - \hat{a} \right]^\top \Sigma \left[a - \hat{a} \right] \right \} \\
    & \times \exp \left \{-\frac{1}{2} \mu^\top \left[ \Phi^{-1} - \Phi^{-1} \Sigma^{-1} \Phi^{-1}\right] \mu + \mu^\top \Phi^{-1} \Sigma^{-1} F^\top N^{-1} \delta t\right\}
\end{aligned}
\label{eq:rearranged_likelihood}
\end{equation}
where we recongnize the Woodbury matrix inversion $[N + F^\top \Phi F]^{-1} = N^{-1} - N^{-1} F \Sigma^{-1} F^\top N^{-1}$. Marginalizing over the coefficients $a$ gives\footnote{The integral over $a$ is a standard Gaussian integral, here producing a term equal to $|2\pi \Sigma^{-1}|^{1/2}$ in the numerator. The $2\pi$ factor simplifies with $|2\pi\Phi|$ in the denominator because $\Sigma$ and $\Phi$ have the same rank.}

\begin{equation}
\begin{aligned}
    \mathcal{L}(\delta t | \Phi, \mu) & = \int_{-\infty} ^{+\infty} da \mathcal{L}(\delta t |a, \Phi, \mu)\\
    & = \frac{\exp \left \{ -\frac{1}{2}\delta t^\top \left[N + F^\top \Phi F \right]^{-1} \delta t \right \}}{|2\pi N|^{1/2} |\Phi|^{1/2} |\Sigma|^{1/2}} \\
    & \times \exp \left \{-\frac{1}{2} \mu^\top \left[ \Phi^{-1} - \Phi^{-1} \Sigma^{-1} \Phi^{-1}\right] \mu + \mu^\top \Phi^{-1} \Sigma^{-1} F^\top N^{-1} \delta t\right\}
\end{aligned}
\end{equation}
where in the denominator we have the Woodbury determinant $|N + F^\top \Phi F| = |N||\Phi||\Sigma|$. Then, the left hand side corresponds to a Gaussian likelihood centered on zero with covariance $C = N + F^\top \Phi F$. This expression generalizes the marginalized likelihood presented in \cite{new_advances_gp} for non central Gaussian priors. For $\mu = 0$, they are identical.

When considering an array of $N_p$ pulsars, the analytical expression of the likelihood is identical, except that now, the data $\delta t$ and basis $F$ are a concatenation of the timing residuals $\delta t_I$ and basis $F_I$ of each pulsar $I$. Additionally, when considering spatially correlated noise between pulsars the Gaussian prior on Fourier coefficients $a$ must incorporate these correlations between $a_I$ of different pulsars and the $\Phi$ is a non-diagonal matrix. As a consequence, the likelihood cannot be factorized in a product of individual pulsar likelihood and the covariance $C$ has a block structure of size $N_p \times N_p$ where each block is $C_{IJ} = N_I \delta_{IJ} + F_I^\top \Phi_{IJ} F_J$. For HD correlated noise, $\Phi_{IJ}\propto \chi^{(2)}_{IJ}$. The inversion of the covariance $C^{-1}=[N + F^\top \Phi F]^{-1}$ then becomes computationally expensive.

\section{Gram-Charlier A expansion for kurtosis}
\label{app:kurtosis_expansion}

We can add the contributions of higher order statistics by expanding the prior probability distribution $\Phi(\vec{a}_n)$ in terms of the Gram-Charlier A series (or similarly using the Edgeworth series \cite{racine_ng}). Then, the non-Gaussian distribution is approximated from the Gaussian prior in \autoref{eq:hd_prior} as

\begin{equation}
    p(\vec{a}_n) \approx \Pi(\vec{a}_n) \times \left[1 + \frac{1}{4!}\sum_{IJKL} \kappa_{IJKL} H_{IJKL}(\vec{a}_n) \right]
\end{equation}
where $\kappa_{IJKL} \propto \chi^{(4)}_{IJKL} h_n^4$ the 4th cumulant (or excess kurtosis), $h_n^4$ the level of excess kurtosis at frequency $n$, and $H_{IJKL}$ the multivariate Hermite polynomials expressed in terms of the $\vec{a}_n$.

This expression has to be normalized and is only valid for small excess kurtosis because this type of expansion behaves poorly for large $h_n^4$ and might even produce unphysical negative probability density values \cite{Contaldi_2000}. Still, it is possible to analytically marginalize the likelihood with respect to the coefficients $\vec{a}_n$ using Isserlis's theorem for different combinations of the elements of $\vec{a}_n$. We have

\begin{equation}
    p(\vec{a}_n) \approx \mathcal{C}^{-1} \times \Pi(\vec{a}_n) \times \left[1 + \frac{1}{4!}\sum_{IJKL} \kappa_{IJKL} H_{IJKL}(\vec{a}_n) \right]
\end{equation}
with $\mathcal{C}$ a normalization coefficient and

\begin{equation}
    \Pi(\vec{a}_n) = \frac{\exp \{-\frac{1}{2}\sum_{IJ} a_{I,n} [\chi_{IJ} \Phi_n]^{-1}a_{J,n}\}}{\sqrt{\det{2\pi\chi \Phi_n}}}.
\end{equation}
where $\chi$ is the matrix of 2-point correlation coefficients $\chi_{IJ}^{(2)}$.

The multivariate $4$-th order Hermite polynomials $H_{IJKL}$ are given by the derivatives of the distribution $\Pi(\vec{a}_n)$ with respect to the components $a_I$ of $\vec{a}_n$ as \cite{kitaura_2012, berkowitz_garner_1970}

\begin{equation}
\begin{aligned}
    H_{IJKL} (\vec{a}) = & (-1)^4 \ \Pi(\vec{a})^{-1} \frac{\partial^4}{\partial a_I \partial a_J \partial a_K \partial a_L} \Pi(\vec{a}) \\
    = & \Phi^{-4} \left[(\sum_{i} \chi_{iI}^{-1}a_{i})(\sum_{j} \chi_{jJ}^{-1}a_{j})(\sum_{k} \chi_{kK}^{-1}a_{k})(\sum_{l} \chi_{lL}^{-1}a_{l}) \right] \\
    & - \Phi^{-3}\left[\chi^{-1}_{IJ} (\sum_{k} \chi_{kK}^{-1}a_{k}) (\sum_{l} \chi_{lL}^{-1}a_{l}) + \chi^{-1}_{IK} (\sum_{j} \chi_{jJ}^{-1}a_{j}) (\sum_{l} \chi_{lL}^{-1}a_{l}) + \chi^{-1}_{IL} (\sum_{j} \chi_{jJ}^{-1}a_{j}) (\sum_{k} \chi_{kK}^{-1}a_{k}) \right.\\
    & - \left.\chi^{-1}_{JK} (\sum_{i} \chi_{iI}^{-1}a_{i}) (\sum_{l} \chi_{lL}^{-1}a_{l}) + \chi^{-1}_{JL} (\sum_{i} \chi_{iI}^{-1}a_{i}) (\sum_{k} \chi_{kK}^{-1}a_{k}) + \chi^{-1}_{KL} (\sum_{i} \chi_{iI}^{-1}a_{i}) (\sum_{j} \chi_{jJ}^{-1}a_{j}) \right]\\
    & + \Phi^{-2} \bigg[\chi^{-1}_{IJ} \chi^{-1}_{KL} + \chi^{-1}_{IK} \chi^{-1}_{JL} + \chi^{-1}_{IL} \chi^{-1}_{JK} \bigg].
\end{aligned}
\end{equation}
where we have dropped the index $n$ for better visibility.

Then, using the rearranged form of the Gaussian likelihood $\mathcal{L}(\delta t |a, \Phi, \mu=0)$ in \autoref{eq:rearranged_likelihood}, the likelihood $\mathcal{L}(\delta t | \Phi, \kappa)$ marginalized with respect to coefficients $a$ is given by

\begin{equation}
\begin{aligned}
    \mathcal{L}(\delta t | \Phi, \kappa) & = \int_{-\infty} ^{+\infty} da \mathcal{L}(\delta t |a, \Phi, \kappa, \mu=0)\\
    & = \mathcal{C}^{-1} \times \mathcal{L}(\delta t |\Phi, \kappa=0) \int da \frac{\exp \left \{ -\frac{1}{2} \left[a - \hat{a} \right]^\top \Sigma \left[a - \hat{a} \right] \right \}}{|2\pi \Sigma^{-1}|}\left[1 + \frac{1}{4!}\sum_{IJKL} \kappa_{IJKL} H_{IJKL}(\vec{a}_n) \right]\\
    & = \mathcal{C}^{-1} \times \mathcal{L}(\delta t|\Phi, \kappa=0) \left[1 + \frac{1}{24} \sum_{IJKL} \kappa_{IJKL} \textrm{E}_\Sigma \left \{H_{IJKL} \right \} \right]
\end{aligned}
\end{equation}
with $\textrm{E}_\Sigma \left \{H_{IJKL} \right \}$ denoting the expected value of the Hermite polynomials for $a \sim \mathcal{N}(\hat{a}, \Sigma)$ that can be obtained using Isserlis's theorem

\begin{equation}
\begin{aligned}
    \textrm{E}_\Sigma \left \{H_{IJKL} \right \} & = \Phi^{-4}\sum_{ijkl} \chi^{-1}_{iI} \chi^{-1}_{jJ} \chi^{-1}_{kK} \chi^{-1}_{lL} \textrm{E}_\Sigma \left \{a_{i} a_{j} a_{k} a_{l}\right \}\\
    & - \Phi^{-3} \left[ \chi^{-1}_{IJ} \sum_{kl}\chi^{-1}_{kK} \chi^{-1}_{lL} \textrm{E}_\Sigma \left \{a_{k} a_{l}\right \} + \textrm{other permutations} \right]\\
    & + \Phi^{-2} \left[ \chi^{-1}_{IJ} \chi^{-1}_{KL} + \textrm{other permutations} \right].
\end{aligned}
\label{eq:hermite_average}
\end{equation}
where

\begin{equation}
\begin{aligned}
    \textrm{E}_\Sigma \left \{a_{i} a_{j} a_{k} a_{l}\right \} & = \hat{a}_i \hat{a}_j \hat{a}_k \hat{a}_l\\
    & + \hat{a}_i \hat{a}_j \Sigma_{kl}^{-1} + \hat{a}_i \hat{a}_k \Sigma_{jl}^{-1} + \hat{a}_i \hat{a}_l \Sigma_{jk}^{-1} + \hat{a}_j \hat{a}_k \Sigma_{il}^{-1} + \hat{a}_j \hat{a}_l \Sigma_{ik}^{-1} + \hat{a}_k \hat{a}_l \Sigma_{ij}^{-1}\\
    & + \Sigma_{ij}^{-1} \Sigma_{kl}^{-1} + \Sigma_{il}^{-1}\Sigma_{jk}^{-1} + \Sigma_{ik}^{-1} \Sigma_{jl}^{-1}
\end{aligned}
\end{equation}
and
\begin{equation}
   \textrm{E}_\Sigma \left \{a_{k} a_{l}\right \} = \Sigma^{-1}_{kl} + \hat{a}_k \hat{a}_l
\end{equation}

The calculation of each component $\Sigma_{IJ}^{-1}$ requires knowing the inverse of matrix $\Sigma$, which is an expensive task. In general, this inverse is implicitly calculated using Cholesky linear system solver to compute $\Sigma^{-1} \delta t$ thus greatly optimizing the computation time. For now, we found no way of optimizing the calculation of \autoref{eq:hermite_average}.

Finally, using \autoref{eq:hermite_average}, we can calculate the normalization coefficient $\mathcal{C}$ for $a \sim \mathcal{N}(0, \chi^{(2)} \Phi_n)$, that is, following the distribution $\Pi(\vec{a}_n)$

\begin{equation}
\begin{aligned}
        \mathcal{C} = 1 + \frac{1}{24} \Phi_n^{-2}\sum_{IJKL} \kappa_{IJKL} \bigg[ & \sum_{ijkl} \chi^{-1}_{iI} \chi^{-1}_{jJ} \chi^{-1}_{kK} \chi^{-1}_{lL}\bigg(\chi_{ij}\chi_{kl} + \chi_{il}\chi_{kj} + \chi_{ik}\chi_{jl}\bigg)\\
        & -\chi^{-1}_{IJ} \left(\sum_{kl} \chi_{kK}^{-1}\chi_{lL}^{-1} \chi_{kl} \right) - \chi^{-1}_{IK} \left(\sum_{jl} \chi_{jJ}^{-1}\chi_{lL}^{-1} \chi_{jl} \right) - \chi^{-1}_{IL} \left(\sum_{jk} \chi_{jJ} \chi_{kK}^{-1} \chi_{jk}\right)\\
        & - \chi^{-1}_{JK} \left(\sum_{il} \chi_{iI}^{-1} \chi_{lL}^{-1} \chi_{il} \right) - \chi^{-1}_{JL} \left(\sum_{ik} \chi_{iI}^{-1} \chi_{kK}^{-1} \chi_{ik} \right) - \chi^{-1}_{KL} \left(\sum_{ij} \chi_{iI}^{-1}\chi_{jJ}^{-1} \chi_{ij} \right)\\
        & +\chi^{-1}_{IJ} \chi^{-1}_{KL} + \chi^{-1}_{IK} \chi^{-1}_{JL} + \chi^{-1}_{IL} \chi^{-1}_{JK} \bigg].
\end{aligned}
\end{equation}

In the case of Edgeworth or Gram-Charlier A expansion, the marginalization with respect to coefficients $a$ yields a quite unpractical expression of the likelihood. Especially for a large array of pulsars, the number of terms in the sum becomes extremely large. If such a method was to be used for PTA data analysis, sampling $a$ from the unmarginalized likelihood might be more efficient and practical.

\section{PP plot}
\label{app:pp}

To validate the simulations, we need to ensure that the data generation and the fitting tool are calibrated. Since the true parameter values $\theta_0$ are known in each simulation, we compute their posterior quantiles $q(\theta_0)$ from the 1d marginalized posterior distributions of each parameter. According to \cite{cook}, if the model and fitting are correct, these quantiles should follow a Uniform(0, 1) distribution. The PP plot in \autoref{fig:pp_plot} seems to show that we indeed recover a uniform distribution within error bars.

\begin{figure}[h!]
\centering
    \includegraphics[width=0.45\textwidth]{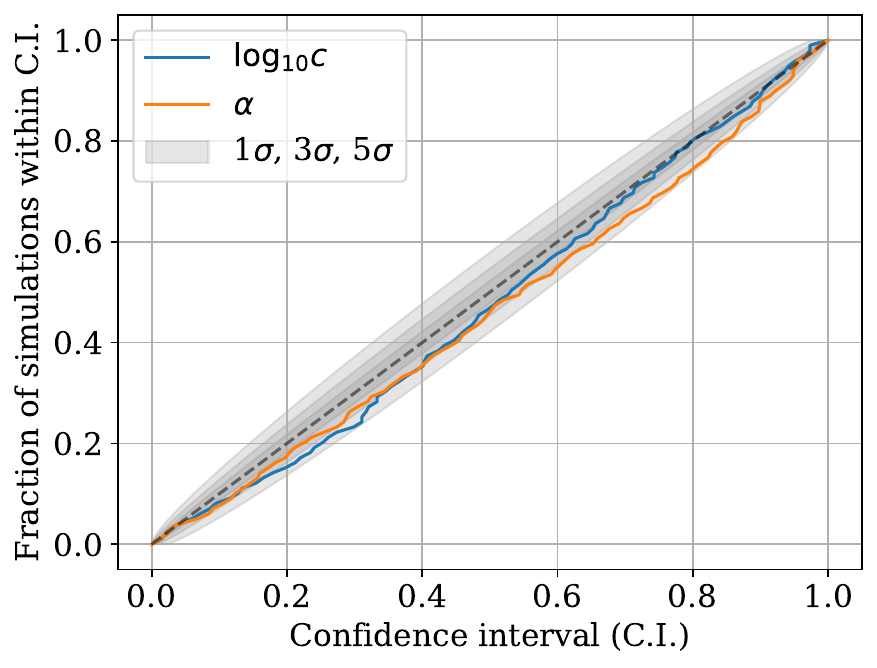}
    \caption{PP plot comparing the fraction of simulations within confidence interval produced from 100 simulations for the non-Gaussianity parameters $c$ and $\alpha$. The grey shaded regions are the 1, 3 and 5-$\sigma$ errors.}
    \label{fig:pp_plot}
\end{figure}

\section{Performance with number of pulsars $N_p$}
\label{app:bf_npsrs}

In this appendix, we show the dependence of the recovered Bayes factor $\mathcal{B}^{NG}_G$ on the number of pulsars in the array $N_p$, justifying our choice of 100 pulsars. The comparison is obtained for the same realization of non-Gaussian noise, only varying the number of pulsars. The conditions of the simulation are the same as those presented in \autoref{sec:results} for different $N_p$.

\begin{figure}[h!]
\centering
    \includegraphics[width=0.45\textwidth]{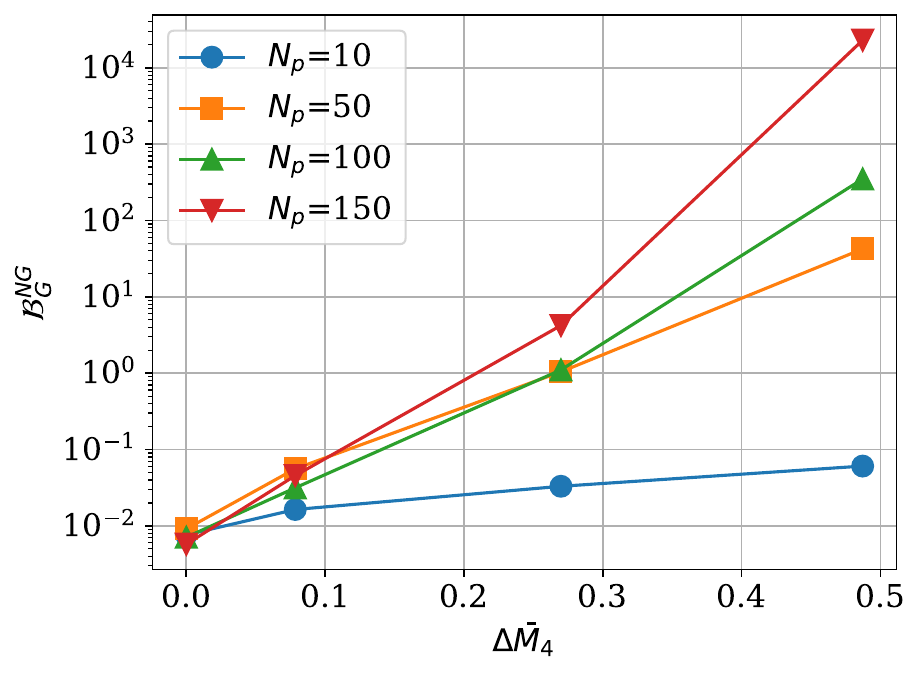}
    \caption{Bayes factor $\mathcal{B}^{NG}_G$ between the non-Gaussian and Gaussian models as a function of the injected relative excess kurtosis $\Delta \bar{M}_4$ calculated using \autoref{eq:excess_moments} for one realization of the noise, with varying number of pulsar $N_p$.}
    \label{fig:bf_npsrs}
\end{figure}

Since this plot is obtained for one realization, it does not show any variance as in \autoref{fig:crn_per_kurt_bf}.
\end{widetext}

\bibliography{ng_pta}

\end{document}